\documentclass[aps,prb,twocolumn,amsmath,%
showpacs,showkeys,floatfix]{revtex4}

\usepackage{epsfig}
\usepackage[english]{babel}
\usepackage{graphicx}
\usepackage{graphics}
\usepackage{amsmath}
\usepackage{url}
\usepackage{amssymb}
\usepackage{subfigure}

\begin{document}

\renewcommand{\figurename}{FIG.}
\newcommand{\vc}[1]{\mathbf{#1}}

\title{The impact of dynamical screening on the phonon dynamics of LaCuO}
\author{Thomas Bauer}
\author{Claus Falter}
\email[Email to: ]{falter@uni-muenster.de}
\affiliation{Institut f\"ur Festk\"orpertheorie, Westf\"alische
Wilhelms-Universit\"at,\\ Wilhelm-Klemm-Str.~10, 48149 M\"unster,
Germany}
\date{\today}

\begin{abstract}
It is shown that dynamical screening of the Coulomb interaction in
LaCuO leads to low-energy electronic collective excitations in a small
region around the $c$-axis, strongly mixing with certain interlayer
phonons polarized along this axis. The manifestation of such a
phonon-plasmon scenario in layered systems based on a nonadiabatic
charge response is quantitatively supported by a realistic calculation
of the frequency and wavevector dependent irreducible polarization part
of the density response function (DRF). The latter is used within
linear response theory to calculate the coupled mode dispersion in the
main symmetry directions of the Brillouin zone (BZ) and the charge
density redistributions excited by certain strongly coupling
phonon-like and plasmon-like modes. Moreover, the corresponding mode
induced orbital averaged changes of the selfconsistent potential felt
by the electrons are assessed. Our analysis should be representative
for the optimally to overdoped state of the cuprates where experimental
evidence of a coherent three-dimensional Fermi surface and a coherent
$c$-axis charge transport is given. It is demonstrated that modes from
the outside of a small region around the $c$-axis can reliably be
calculated within the adiabatic limit. Only a minor nonadiabatic
correction is found for these modes in form of stiffening of the
high-frequency oxygen-bond-stretching mode at the $X$-point which is
attributed to dynamical reduced nesting. On the other hand, modes
inside the nonadiabatic sector of the BZ have to be determined
nonadiabatically owing to the poor dynamical screening of the
long-ranged Coulomb interaction around the $c$-axis by the slow charge
dynamics. In particular, the relevance of the strongly coupling
phonon-like apex oxygen $Z$-point breathing mode at about 40 meV is
emphasized whose mode energy decreases with less doping. Comparing the
calculations with experimental neutron scattering and infrared
measurements provides strong evidence for low-lying plasmons around the
$c$-axis coupling strongly with corresponding optical phonons. The
existence of these coupled modes evidenced by the calculations yields
additional virtual exchange bosons for pairing. Finally, the strong
nonlocal, nonadiabatic polar electron-phonon coupling around the
$c$-axis found in the computations means that besides the repulsive
short-ranged part of the Coulomb interaction responsible for strong
correlation effects also the long-ranged part plays an important role
for the physics in the cuprates.
\end{abstract}



\maketitle

\section{Introduction}\label{SecOne}
In layered, strongly anisotropic materials like the copper oxide
high-temperature superconductors (HTSC's) the screening of the Coulomb
interaction between the layers is imperfect and its dynamic nature
becomes important. Perpendicular to the CuO planes low-lying electronic
collective modes (interlayer plasmons) can be expected. In connection
with superconductivity the interplay between the attractive
interaction, mediated e.g. by phonons, and the dynamically screened
Coulomb interaction is more complex than in the conventional theory.
Accordingly, the existence and possible relevance of energetically low
lying plasmons, e.g. accoustic plasmons, for superconductivity in the
cuprates has been discussed before in the literature, see e.g. Refs.
\onlinecite{Ref01,Ref02,Ref03,Ref04,Ref05,Ref06}. A direct observation
of low lying plasmons in the HTSC's, e.g. by high-resolution inelastic
electron scattering or some other technique is still missing. As shown
in this work there is only a small $\vc{q}$-space region around the
$c$-axis with a nonadiabatic charge response where low lying
phonon-plasmon modes can exist and thus a very high $\vc{q}$-space
resolution would be needed in an experiment. However, there is indirect
experimental evidence from different probes, like the use of
reflectance and ellipsometric measurements \cite{Ref07}, external
losses in photoemission \cite{Ref08}, infrared reflectivity
\cite{Ref09} or resonant inelastic $x$-ray scattering \cite{Ref10}.

A proper account of dynamical screening of the Coulomb interaction is
not only substantial for superconductivity, but also for a correct
description of normal state properties of the HTSC's such as phonon
dynamics, see our simplified model approach in Refs.
\onlinecite{Ref11,Ref12} and references therein. These calculations
indicate that in a small sector around the $c$-axis of LaCuO
phonon-plasmon mixing becomes likely if the interlayer coupling is
sufficiently weak. The reason for such a mixing behaviour is a
sufficiently slow electron dynamics perpendicular to the CuO planes.
However, the important question remains if this is really true in LaCuO
or other cuprates. This topic and its consequences is investigated in
the present work. Electron dynamics and phonon dynamics then have about
the same time scale and as a consequence nonadiabatic behaviour
results.

Due to the incomplete dynamical screening of the Coulomb interaction a
strong nonlocal, polar electron-phonon coupling results along the
$c$-axis together with optical conductivity, also in the well-doped
metallic state. Experimentally this can be seen for example by the
infrared measurements for LaCuO \cite{Ref13,Ref14,Ref15,Ref16}. Direct
evidence for the presence and importance of this type of unconventional
electron-phonon coupling around the $c$-axis where the electrons in the
cuprates are strongly coupled to $c$-axis polar phonons also comes from
a recent study of the self-energy of the nodal quasiparticles in Bi2201
(Ref. \onlinecite{Ref17}). For a discussion of phonon-plasmon mixing in
the cuprates in the context of many-body polaronic effects in the
phonon spectrum, we refer to Ref. \onlinecite{Ref18}.

As far as underdoped and even optimally doped LaCuO is concerned, the
optical $c$-axis spectra display the features typical for an ionic
insulator \cite{Ref13,Ref14,Ref15,Ref16} and not those of a metal dealt
within adiabatic approximation. The spectra are dominated by optical
phonons and are almost unchanged from that of the insulating parent
compound upon doping. These experimental facts cannot be explained by
the first principles calculations of the phonon dynamics of the HTSC's
published so far in the literature \cite{Ref19,Ref20,Ref21,Ref22}
because these computations have been performed within (static) density
functional theory (DFT) mostly in local density approximation (LDA).
Such calculations are based on the adiabatic approximation and always
yield a static metallic screening behaviour consistent with closed
LO-TO splittings, at the $\Gamma$-point, i.e.  in particular closed
$A_{2u}$-splittings in LaCuO. This is clearly in contrast to the
optical activity in the infrared experiments. In static DFT
calculations the transverse effective charges vanish in the metallic
state and consequently the induced dipole moments defining the
oscillator strenghts in the dielectric function (matrix) vanish, too.
So, there will be no optical activity by the phonons.

In the first principles calculations \cite{Ref19,Ref20,Ref21,Ref22}
also no explicit results for the phonon dispersion curves along the
$c$-axis or a small region around this axis have been presented.
However, this has been accomplished within a simplified model in Refs.
\onlinecite{Ref11,Ref12} and will quantivatively be computed in this
work. In Ref. \onlinecite{Ref20} only a short remark is given for the
case of LaCuO stating that certain experimental results along the
$c$-axis are not supported by the calculation. Moreover, when judging
the findings in Refs. \onlinecite{Ref22,Ref23} of a small phonon
contribution to the photoemission kink detected by ARPES
\cite{Ref24,Ref25} one should keep in mind that these very interesting
calculations also have been performed within static DFT, neglecting
e.g. the strong nonadiabatic dynamically screened polar coupling around
the $c$-axis and also correlation effects beyond LDA. On the other
hand, the importance of the electron-phonon interaction in context with
the kink in the dispersion of the nodal quasiparticles is emphasized in
Ref. \onlinecite{Ref26} on the basis of the two-dimensional three-band
Hubbard model with electron-phonon interaction included and also in
Ref. \onlinecite{Ref27}.

Because of the strong anisotropy of the cuprates and consequently an
expected very weak dispersion of the electronic  bandstructure along
the $c$-axis, which, however, is systematically overestimated by
DFT-LDA calculations,  electron dynamics and $c$-axis phonon dynamics
is coupled dynamically and a nonadiabatic treatment with a dynamical
screened Coulomb interaction is essential. On the other hand, first
principles calculations \cite{Ref19,Ref20,Ref21,Ref22} and also our
microscopic calculations, see e.g. Refs.
\onlinecite{Ref12,Ref28,Ref29}, have shown that the static DFT is in
general a sufficient approximation for phonon modes propagating in the
CuO-plane  of the HTSC's, but probably not so around the $c$-axis.

For modes from this $\vc{q}$-space region a coupled phonon-plasmon
scenario has been suggested for LaCuO within a simplified microscopic
model approach in the framework of linear response theory
\cite{Ref11,Ref30} assuming a sufficiently weak interlayer coupling. In
this treatment an eleven-band model (11BM) that represents the
two-dimensional electronic structure of the CuO plane in terms of the
Cu3d and O$_{xy}$2p orbitals is generalized to the third dimension by
introducing a suitable interlayer coupling in parametrized form. By
varying this coupling from the outside it is possible  to pass from the
strictly two-dimensional nonadiabatic case of the charge response
(charge confinement in the CuO plane) to a moderate anisotropic
adiabatic situation, typical for DFT-LDA calculations. A comparison
with experimental results from inelastic neutron scattering (INS) and
infrared spectroscopy also has been given. From the interpretation of
the results a phonon-plasmon mixing seems to be likely. Adopting within
such a model approach a suitable coupling we were able to explain at
least qualitatively an appearent inconsistency between the current
interpretation of the INS results for the $\Lambda_{1} \sim (0,0,1)$
branches in doped metallic samples of LaCuO (Refs.
\onlinecite{Ref31,Ref32}). The branches look as expected in adiabatic
approximation for a less anisotropic metal, i.e. featuring closed
$A_{2u}$ splittings. On the other hand, the infrared data display as
already mentioned significant $A_{2u}$-splittings typical for an ionic
insulator, even for well doped metallic probes. Moreover, in Ref.
\onlinecite{Ref11} arguments have been presented in the framework of
phonon-plasmon mixing to understand at least qualitatively the large
softening and the massive line broadening of the apical oxygen
breathing mode O$_z^Z$, at the $Z$-point of the BZ. Altogether, the
model calculations  point to phonon-plasmon coupling in a small
nonadiabatic region around the $c$-axis where the Coulomb interaction
has to be screened dynamically. A direct experimental search for such a
scenario is inconclusive so far \cite{Ref32,Ref33}.

From our calculations in this work we can conclude that much better
$\vc{q}$-space resolution transverse to the $c$-axis would be needed in
the experiment to resolve the coupled mode dispersion in the small
nonadiabatic sector.

One important topic of this paper is to scrutinize and quantify the
discussion performed within the simplified model for LaCuO by starting
with a realistic tight binding approximation (TBA) of the first
principles bandstructure (31-band-model (31BM), Ref.
\onlinecite{Ref34}) which is as a typical LDA-like bandstructure not
anisotropic enough and adapt the latter to the real anisotropy of the
material. With such a realistic electronic bandstructure the wavevector
and frequency-dependent irreducible polarization part of the DRF is
calculated. The latter is used in linear response theory to obtain the
dynamically screened Coulomb interaction and ultimately the full
nonadiabatic coupled mode dispersion in the main symmetry directions of
the BZ. Moreover, charge density redistributions $\delta \rho$ of
certain strongly coupling phonon-like and plasmon-like modes of the
nonadiabatic region and the corresponding mode induced changes of the
selfconsistent potential $\delta V$ felt by the electrons are
investigated and compared with the adiabatic limit. From our
calculations a phonon-plasmon mixing and a strong nonlocal,
nonadiabatic polar electron-phonon interaction is predicted for LaCuO.
This conclusion should be representative for the optimally to overdoped
state of the cuprates where experimental evidence of a coherent
three-dimensional Fermi surface (FS) \cite{Ref35} and coherent $c$-axis
charge transport is given \cite{Ref36,Ref37,Ref38}.

The article is organized as follows. In Sec. II some elements of the
theory necessary to understand the calculated results are reviewed.
Section III presents the computations. It gives a discussion of the
electronic structure and the nonadiabatic  and adiabatic mode
dispersion. Furthermore, $\delta \rho$ and $\delta V$ are investigated
for certain relevant generic modes in the cuprates. Finally, a summary
of the paper is presented in Sec. IV and the conclusions are drawn.

\section{ELEMENTS OF THE THEORY AND MODELING}\label{SecTwo}
In the following a survey of the theory and modeling is presented. A
detailed description can be found in Ref. \onlinecite{Ref39} and in
particular in Ref. \onlinecite{Ref40} where the calculation of the
coupling parameters of the theory is presented.

The local part of the electronic charge response and the EPI is
approximated in the spirit of the quasi-ion approach \cite{Ref29,Ref41}
by an ab initio rigid ion model (RIM) taking into account covalent ion
softening in terms of (static) effective ionic charges calculated from
a tight-binding analysis. The tight-binding analysis supplies these
charges as extracted from the orbital occupation numbers $Q_{\mu}$ of
the $\mu$ (tight-binding) orbital in question:
\begin{equation}\label{Eq1}
Q_{\mu} = \frac{2}{N} \sum\limits_{n\vc{k}} |C_{\mu n} (\vc{k})|^2.
\end{equation}
$C_{\mu n} (\vc{k})$ stands for the $\mu$-component of the eigenvector
of band $n$ at the wavevector $\vc{k}$ in the first BZ; the summation
in (\ref{Eq1}) runs over all occupied states and $N$ gives the number
of the elementary cells in the (periodic) crystal.

In addition, scaling of the short-ranged part of certain pair
potentials between the ions is performed to simulate further covalence
effects in the calculation in such a way that the energy-minimized
structure is as close as possible to the experimental one \cite{Ref42}.
Structure optimization and energy minimization is very important for a
reliable calculation of the phonon dynamics through the dynamical
matrix. Taking just the experimental structure data as is done in many
cases in the literature may lead to uncontrolled errors in the phonon
calculations.

The RIM with the corrections just mentioned then serves as an unbiased
reference system for the description of the HTSC's and can be
considered as a first approximation for the insulating state of these
compounds.  Starting with such an unprejudiced rigid reference system
non-rigid electronic polarization processes are introduced in form of
more or less localized electronic charge-fluctuations (CF's) at the
outer shells of the ions. Especially in the metallic state of the
HTSC's the latter dominate the {\em nonlocal} contribution of the
electronic density response and the EPI and are particularly important
in the CuO planes. In addition, \textit{anisotropic}
dipole-fluctuations (DF's) are admitted in our approach
\cite{Ref40,Ref43}, which prove to be specifically of interest for the
ions in the ionic layers mediating the dielectric coupling and for the
polar modes. Thus, the basic variable of our model is the ionic density
which is given in the perturbed state by
\begin{equation}\label{Eq2}
\rho_\alpha(\vc{r},Q_\lambda, \vc{p}_\alpha) =
\rho_\alpha^0(r) + \sum_{\lambda}Q_\lambda \rho_\lambda^\text{CF}(r)
+ \vc{p}_\alpha \cdot
\hat{\vc{r}} \rho_\alpha^\text{D}(r).
\end{equation}
$\rho_\alpha^0$ is the density of the unperturbed ion, as used in the
RIM, localized at the sublattice $\alpha$ of the crystal and moving
rigidly with the latter under displacement. The $Q_\lambda$ and
$\rho^\text{CF}_\lambda$ describe the amplitudes and the form-factors
of the CF's and the last term in Eq. \eqref{Eq2} represents the dipolar
deformation of an ion $\alpha$ with amplitude (dipole moment)
$\vc{p}_\alpha$ and a radial density distribution
$\rho_\alpha^\text{D}$.  $\hat{\vc{r}}$ denotes the unit vector in the
direction of $\vc{r}$. The $\rho^\text{CF}_\lambda$ are approximated by
a spherical average of the orbital densities of the ionic shells
calculated in LDA taking self-interaction effects (SIC) into account.
The dipole density $\rho_\alpha^\text{D}$ is obtained from a modified
Sternheimer method in the framework of LDA-SIC \cite{Ref40}. All
SIC-calculations are performed for the average spherical shell in the
orbital-averaged form according to Perdew and Zunger \cite{Ref44}. For
the correlation part of the energy per electron $\epsilon$ the
parametrization given in Ref. \onlinecite{Ref44} has been used.

The total energy of the crystal is obtained by assuming that the
density can be approximated by a superposition of overlapping densities
$\rho_\alpha$. The $\rho_\alpha^0$ in Eq. \eqref{Eq2} are also
calculated within LDA-SIC taking environment effects, via a Watson
sphere potential and the calculated static effective charges of the
ions into account. The Watson sphere method is only used for the oxygen
ions and the depth of the Watson sphere potential is set as the
Madelung potential at the corresponding site. Such an  approximation
holds well in the HTSC's \cite{Ref42,Ref45}. As a general rule, partial
covalence reduces the amplitude of the static effective charges in
mixed ionic-covalent compounds like the HTSC's, because the charge
transfer from the cations to the anions is not complete as in the
entirely ionic case. Finally, applying the pair-potential approximation
we get for the total energy:
\begin{equation}\label{Eq3}
E(R,\zeta) = \sum_{\vc{a},\alpha} E_\alpha^\vc{a}(\zeta)
+\frac{1}{2}\sum_{(\vc{a},\alpha)\neq(\vc{b},\beta)}
\Phi_{\alpha\beta}
\left(\vc{R}^\vc{b}_\beta-\vc{R}^\vc{a}_\alpha,\zeta\right).
\end{equation}
The energy $E$ depends on both the configuration of the ions $\{R\}$
and the electronic (charge) degrees of freedom (EDF) $\{\zeta\}$ of the
charge density, i.e. $\{Q_\lambda\}$ and $\{\vc{p}_\alpha\}$ in Eq.
\eqref{Eq2}.  $E_\alpha^\vc{a}$ are the energies of the single ions.
$\vc{a}$, $\vc{b}$ denote the elementary cells and $\alpha$, $\beta$
the corresponding sublattices.  The second term in Eq. \eqref{Eq3} is
the interaction energy of the system, expressed in terms of
\textit{anisotropic} pair-interactions $\Phi_{\alpha\beta}$. Both
$E_\alpha^\vc{a}$ and $\Phi_{\alpha\beta}$ in general depend upon
$\zeta$ via $\rho_\alpha$ in Eq. \eqref{Eq2}.

The pair potentials in Eq. \eqref{Eq3} can be seperated into
long-ranged Coulomb contributions and short-ranged terms as follows:
\begin{align}\nonumber
\Phi_{\alpha\beta}(\vc{R},\zeta) =& \frac{\mathcal{Z}_\alpha \mathcal{Z}_\beta}{R}
-(\mathcal{Z}_\alpha \vc{p}_\beta + \mathcal{Z}_\beta \vc{p}_\alpha)\cdot\frac{\vc{R}}{R^3}
+\frac{\vc{p}_\alpha\cdot\vc{p}_\beta}{R^3}\\\label{Eq4}
&-3\frac{(\vc{p}_\alpha\cdot\vc{R})(\vc{R}\cdot\vc{p}_\beta)}{R^5}
+ \widetilde{\Phi}_{\alpha\beta}(\vc{R},\zeta),
\end{align}

\begin{align}\nonumber
\widetilde{\Phi}_{\alpha\beta}(\vc{R},\zeta) =& K_\alpha U_\beta(\vc{R},\zeta)
+ K_\beta U_\alpha(\vc{R},\zeta)\\\label{Eq5}
&+ W_{\alpha\beta}(\vc{R},\zeta) + G_{\alpha\beta}(\vc{R},\zeta).
\end{align}
The first term in Eq. \eqref{Eq4} describes the long-ranged ion-ion,
the second the dipole-ion and the third and fourth term the
dipole-dipole interaction. $\mathcal{Z}_\alpha$ and $\mathcal{Z}_\beta$
are the variable charges of the ions in case the CF's are excited. The
latter reduce to the ionic charges for rigid ions. $K_\alpha$ and
$K_\beta$ are the charges of the ion cores. The remaining term in Eq.
\eqref{Eq4} given in \eqref{Eq5} represents the short-ranged
interactions. These short-ranged contributions to the pair potentials
are expressed by the following integrals:
\begin{align}\label{Eq6}
U_\alpha(\vc{R},\zeta) &= - \int d^3r \rho_\alpha(\vc{r},\zeta) \times\nonumber\\
&\times \left( \frac{1}{|\vc{r}-\vc{R}|} - \frac{1}{R} - \frac{\vc{r}\cdot\vc{R}}{R^3} \right),
\end{align}
\begin{align} \label{Eq7}
W_{\alpha\beta}(\vc{R},\zeta) &= \int d^3r \int d^3r'
\bigl[\rho_\alpha(\vc{r},\zeta) \rho_\beta(\vc{r}',\zeta) \times\nonumber\\
&\times\left(\frac{1}{|\vc{r}-\vc{r}'-\vc{R}|}-\frac{1}{R}-\frac{(\vc{r}+\vc{r}')\cdot\vc{R}}{R^3} \right) \bigr],
\end{align}
\begin{align} \label{Eq8}
G_{\alpha\beta}(\vc{R},\zeta) &= \int  d^3r \bigl[\rho_{\alpha\beta}(\vc{r},\zeta)\epsilon(\rho_{\alpha\beta}(\vc{r},\zeta))\nonumber\\
&-\rho_{\alpha}(\vc{r},\zeta)\epsilon(\rho_{\alpha}(\vc{r},\zeta))\\
&-\rho_{\beta}(\vc{r}-\vc{R},\zeta)\epsilon(\rho_{\beta}(\vc{r}-\vc{R},\zeta))\bigr],\nonumber
\end{align}
with
\begin{equation}\label{Eq9}
\rho_{\alpha\beta}(\vc{r},\zeta) = \rho_\alpha(\vc{r},\zeta)+\rho_\beta(\vc{r}-\vc{R},\zeta).
\end{equation}
$K_\alpha U_\beta(\vc{R},\zeta)$ yields the short-ranged contribution
of the interaction between the core $\alpha$ and the density
$\rho_\beta$ according to Eq. \eqref{Eq2}.
$W_{\alpha\beta}(\vc{R},\zeta)$ represents the short-ranged Coulomb
contribution of the interaction of the density $\rho_\alpha$ with the
density $\rho_\beta$. $G_{\alpha\beta}(\vc{R},\zeta)$ is the sum of the
kinetic one-particle- and the exchange-correlation (XC) contribution of
the interaction between the two ions \cite{Ref40}. The short-ranged
part of the potentials and the various coupling coefficients are
calculated numerically for a set of disctances $R$ between the ions.
The corresponding results are than described by an analytical function
of the form:
\begin{equation}\label{Eq10}
f(R) = \pm \text{exp}\left(\alpha+\beta R+\frac{\gamma}{R}\right).
\end{equation}
$\alpha$, $\beta$ and $\gamma$ in Eq. \eqref{Eq10} are fit parameters.
From the adiabatic condition
\begin{equation}\label{Eq11}
\frac{\partial E(R,\zeta)}{\partial \zeta} = 0
\end{equation}
an expression for the atomic force constants, and accordingly the
dynamical matrix in harmonic approximation can be derived:
\begin{align}\label{Eq12}\nonumber
t_{ij}^{\alpha\beta}(\vc{q}) &=
\left[t_{ij}^{\alpha\beta}(\vc{q})\right]_\text{RIM}\\ &-
\frac{1}{\sqrt{M_\alpha M_\beta}} \sum_{\kappa,\kappa'}
\left[B^{\kappa\alpha}_i(\vc{q}) \right]^{*} \left[C^{-1}(\vc{q})
\right]_{\kappa\kappa'} B^{\kappa'\beta}_j(\vc{q}).
\end{align}
The first term on the right hand side denotes the contribution from the
RIM. $M_\alpha$, $M_\beta$ are the masses of the ions and $\vc{q}$ is a
wave vector from the first BZ.

The quantities $\vc{B}(\vc{q})$ and $C(\vc{q})$ in Eq. \eqref{Eq12}
represent the Fourier transforms of the electronic coupling
coefficients as calculated from the energy in Eq. \eqref{Eq3}, or the
pair potentials in Eqs. \eqref{Eq4}-\eqref{Eq9}, respectivly.
\begin{align}\label{Eq13}
\vc{B}_{\kappa\beta}^{\vc{a}\vc{b}} &= \frac{\partial^2
E(R,\zeta)}{\partial \zeta_\kappa^\vc{a} \partial R_\beta^\vc{b}},
\\\label{Eq14} C_{\kappa\kappa'}^{\vc{a}\vc{b}} &= \frac{\partial^2
E(R,\zeta)}{\partial \zeta_\kappa^\vc{a} \partial
\zeta_{\kappa'}^\vc{b}}.
\end{align}
$\kappa$ denotes the EDF (CF and DF in the present model, see Eq.
\eqref{Eq2}) in an elementary cell.  The $\vc{B}$ coefficients describe
the coupling between the EDF and the displaced ions (bare
electron-phonon coupling), and the coefficients $C$ determine the
interaction between the EDF. The phonon frequencies
$\omega_\sigma(\vc{q})$ and the corresponding eigenvectors
$\vc{e}^\alpha(\vc{q}\sigma)$ of the modes $(\vc{q}\sigma)$ are
obtained from the secular equation for the dynamical matrix in Eq.
\eqref{Eq12}, i.e.
\begin{equation}\label{Eq15}
\sum_{\beta,j} t_{ij}^{\alpha\beta}(\vc{q})e_j^\beta(\vc{q}) =
\omega^2(\vc{q}) e_i^\alpha(\vc{q}).
\end{equation}
The Eqs. \eqref{Eq12}-\eqref{Eq15} are generally valid and, in
particular, are independent of the specific model for the decomposition
of the perturbed density in Eq. \eqref{Eq2} and the pair approximation
Eq. \eqref{Eq3} for the energy. The lenghty details of the calculation
of the coupling coefficients $\vc{B}$ and $C$ cannot be reviewed in
this paper. They are presented in Ref. \onlinecite{Ref40}. In this
context we remark that the coupling matrix $C_{\kappa\kappa'}(\vc{q})$
of the EDF-EDF interaction, whose inverse appears in Eq. \eqref{Eq12}
for the dynamical matrix, can be written in matrix notation as
\begin{equation}\label{Eq16}
C = \Pi^{-1} + \widetilde{V}.
\end{equation}
$\Pi^{-1}$ is the inverse of the {\em irreducible polarization part} of
the density response function (matrix) and contains the kinetic part to
the interaction $C$ while $\widetilde{V}$ embodies the Hartree and
exchange-correlation contribution. $C^{-1}$ needed for the dynamical
matrix and the EPI is closely related to the (linear) density response
function (matrix) and to the inverse dielectric function (matrix)
$\varepsilon^{-1}$, respectively.

Only very few attempts have been made to calculate the phonon
dispersion and the EPI of the HTSC's using the linear response method
in form of density functional perturbation theory (DFPT) within LDA
\cite{Ref19,Ref20,Ref21}. These calculations correspond to calculating
$\Pi$ and $\widetilde{V}$ in DFT-LDA and for the \textit{metallic}
state only. On the other hand, in our microscopic modeling DFT-LDA-SIC
calculations are performed for the various densities in Eq. \eqref{Eq2}
in order to obtain the coupling coefficients $\vc{B}$ and
$\widetilde{V}$. Including SIC is particularly important for localized
orbitals like Cu3d in the HTSC's. SIC as a correction for a single
particle term is not a correlation effect, which per definition cannot
be described in a single particle theory, but SIC is important for
contracting in particular the localized Cu3d orbitals. Our theoretical
results for the phonon dispersion \cite{Ref12,Ref28,Ref43}, which
compare well with the experiments, demonstrate that the approximative
calculation of the coupling coefficients in our approach is sufficient,
even for the localized Cu3d states. Written in matrix notation we get
for the density response matrix the relation
\begin{equation}\label{Eq17}
C^{-1} = \Pi(1+\widetilde{V}\Pi)^{-1} \equiv \Pi \varepsilon^{-1},
\hspace{.7cm} \varepsilon = 1 + \widetilde{V}\Pi.
\end{equation}
The CF-CF submatrix of the matrix $\Pi$ can approximatively be
calculated for the metallic (but not for the undoped and underdoped)
state of the HTSC's from a TBA of a single particle electronic
bandstructure. In this case the electronic polarizability $\Pi$ in
tight-binding representation reads:
\begin{align}\nonumber
\Pi_{\kappa\kappa'}&(\vc{q},\omega=0) = -\frac{2}{N}\sum_{n,n',\vc{k}}
\frac{f_{n'}(\vc{k}+\vc{q})
-f_{n}(\vc{k})}{E_{n'}(\vc{k}+\vc{q})-E_{n}(\vc{k})}
\times \\\label{Eq18} &\times \left[C_{\kappa n}^{*}(\vc{k})C_{\kappa
n'}(\vc{k}+\vc{q}) \right] \left[C_{\kappa' n}^{*}(\vc{k})C_{\kappa'
n'}(\vc{k}+\vc{q}) \right]^{*}.
\end{align}
$f$, $E$ and $C$ in Eq. \eqref{Eq18} are the occupation numbers, the
single-particle energies and the expansion coefficientes of the
Bloch-functions in terms of tight-binding functions.

The self-consistent change of an EDF at an ion induced by a phonon mode
$(\vc{q} \sigma)$ with frequency $\omega_\sigma(\vc{q})$ and
eigenvector $\vc{e}^\alpha(\vc{q}\sigma)$ can be derived in the form
\begin{eqnarray}\label{Eq19}
\delta\zeta_\kappa^\vc{a}(\vc{q}\sigma) & = & \left[-\sum_\alpha
\vc{X}^{\kappa\alpha}(\vc{q})\vc{u}_\alpha(\vc{q}\sigma)\right]
e^{i\vc{q}\vc{R}_\kappa^\vc{a}}\nonumber\\
& \equiv & \delta\zeta_\kappa(\vc{q}\sigma)e^{i\vc{q}\vc{R}^\vc{a}},
\end{eqnarray}
with the displacement of the ions
\begin{eqnarray}\label{Eq20}
\vc{u}_\alpha^{\vc{a}}(\vc{q}\sigma) & = &
\left(\frac{\hbar}{2M_\alpha\omega_\sigma(\vc{q})}
\right)^{1/2}\vc{e}^\alpha(\vc{q}\sigma)e^{i\vc{q}\vc{R}^\vc{a}}\nonumber\\
& \equiv & \vc{u}_\alpha(\vc{q}\sigma)e^{i\vc{q}\vc{R}^\vc{a}}.
\end{eqnarray}
The self-consistent response per unit displacement of the EDF in Eq.
\eqref{Eq19} is calculated in linear response theory as:
\begin{equation}\label{Eq21}
\vc{X}(\vc{q}) = \Pi(\vc{q})\varepsilon^{-1}(\vc{q})\vc{B}(\vc{q}) =
C^{-1}(\vc{q})\vc{B}(\vc{q}).
\end{equation}
A measure of the strength of the EPI for a certain phonon mode
$(\vc{q}\sigma)$ is provided by the change of the self-consistent
potential in the crystal felt by an electron at some space point
$\vc{r}$, i.e. $\delta V_\text{eff}(\vc{r},\vc{q}\sigma)$.  Averaging
this quantity with the corresponding density form factor
$\rho_\kappa(\vc{r}-\vc{R}_\kappa^\vc{a})$ at the EDF located at
$\vc{R}_\kappa^\vc{a}$, we obtain
\begin{equation}\label{Eq22}
\delta V_\kappa^\vc{a}(\vc{q}\sigma) = \int dV
\rho_\kappa(\vc{r}-\vc{R}_\kappa^\vc{a}) \delta
V_\text{eff}(\vc{r},\vc{q}\sigma)
\end{equation}
as a parameter for the strength of the EPI in the mode $(\vc{q}\sigma)$
mediated by the EDF considered. For an expression of $\delta
V_\kappa^\vc{a}(\vc{q}\sigma)$ in terms of the coupling coefficients in
Eqs. \eqref{Eq13} and \eqref{Eq14}, see Ref. \onlinecite{Ref12}.

The generalization for the quantity $\Pi$ in Eqs. \eqref{Eq16} and
\eqref{Eq17} needed for the kinetic part of the charge response in the
nonadiabatic regime, where dynamical screening effects must be
considered, can be achieved by adding $(\hbar\omega+i \eta)$ to the
differences of the single-particle energies in the denominator of the
expression for $\Pi$ in Eq. \eqref{Eq18}. Other possible nonadiabatic
contributions to $C$ related to dynamical exchange-correlation effects
and the phonons themselves are beyond the scope of the present model.
Using Eq. \eqref{Eq17} for the dielectric matrix, $\varepsilon$, and
the frequency-dependent version of the irreducible polarization part,
$\Pi$, according to Eq. \eqref{Eq18}, the free-plasmon dispersion is
obtained from the condition,
\begin{equation} \label{Eq23}
\text{det} [\varepsilon_{\kappa\kappa'} (\vc{q}, \omega)] = 0 .
\end{equation}
The coupled-mode frequencies of the phonons and the plasmons must be
determined self-consistently from the secular equation \eqref{Eq15} for
the dynamical matrix which now contains the frequency $\omega$
implicitly via $\Pi$ in the response function $C^{-1}$. Analogously,
the dependence on the frequency is transferred to the quantity $\vc{X}$
in Eq. \eqref{Eq21} and thus to $\delta\zeta_\kappa$ and $\delta
V_\kappa$ in Eqs. \eqref{Eq19} and \eqref{Eq22}, respectively. Such a
nonadiabatic approach is necessary for a description of the interlayer
phonons and the charge-response within a small region around the
$c$-axis.

The time-consuming numerical calculations were carried out on the
computers of the Morfeus GRID at the Westf\"alische
Wilhelms-Universit\"at M\"unster, with the use of Condor\cite{Ref100}.

\section{RESULTS AND DISCUSSION} \label{SecThree}
\subsection{Model of the single particle content of the irreducible polarization part}
As a first approximation the electronic bandstructure (BS) $E_{n}
(\vc{k})$ of LaCuO and the expansion coefficients $C_{\kappa n}
(\vc{k})$, needed as input for the single particle content of the
irreducible polarization part $\Pi_{\kappa\kappa'}$ in Eq. \eqref{Eq18}
are taken from an accurate tight-binding representation of the
first-principles linearized-augmented-plane-wave bandstructure (LAPW)
as obtained within the framework of DFT-LDA \cite{Ref34}.

%%%%%%%%%%%%%%%%%%%%%%%%%%%%%%%%%%%%%%%%%%%%%%%%%%%%%%%%%%%%%%%%%%%%%%%%%%%%%%%%%%%%%%%%%
%%%%%%%%%%%%%%%%%%%%%%%%%%%%%%%%%%%%%%%%%%%%%%%%%%%%%%%%%%%%%%%%%%%%%%%%%%%%%%%%%%%%%%%%%
\begin{figure}
\includegraphics[width=\linewidth]{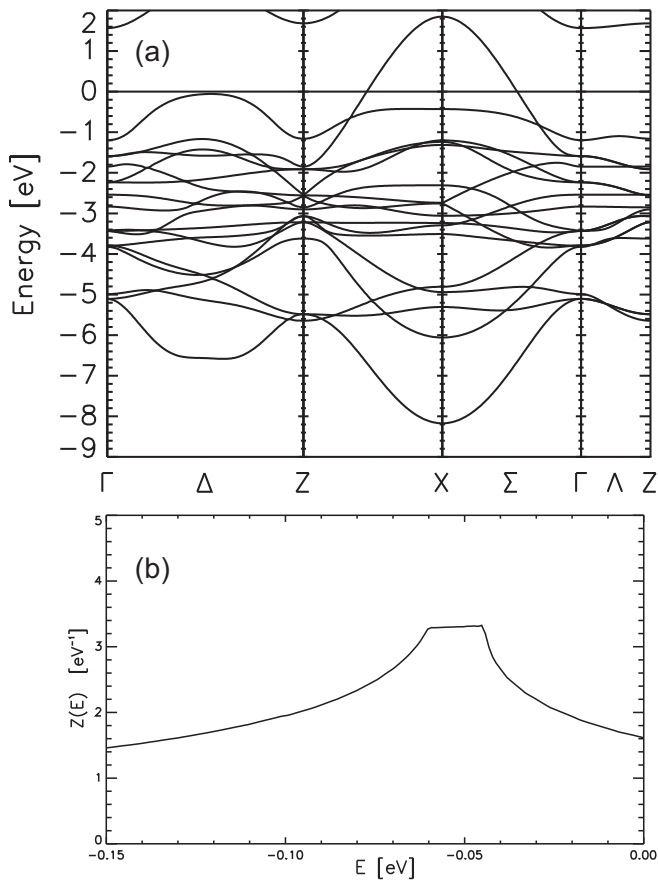}
\caption{Electronic bandstructure $E_{n}(\vc{k})$ of LaCuO in the 31
band model (31BM) obtained from an accurate tight-binding
representation of the first-principles linearized-augmented -plane-wave
bandstructure \cite{Ref34}. The Fermi energy $E_{F}$ corresponding to
the undoped material is taken as the zero of the energy (a). Density of
states $Z(E)$ of the 31BM (b).}\label{fig01}
\end{figure}
%%%%%%%%%%%%%%%%%%%%%%%%%%%%%%%%%%%%%%%%%%%%%%%%%%%%%%%%%%%%%%%%%%%%%%%%%%%%%%%%%%%%%%%%%
%%%%%%%%%%%%%%%%%%%%%%%%%%%%%%%%%%%%%%%%%%%%%%%%%%%%%%%%%%%%%%%%%%%%%%%%%%%%%%%%%%%%%%%%%

This analysis leads to a 31BM including the La5d, Cu3d, 4s, 4p, and O2p
states. The result for the recalculated tight-binding bandstructure is
shown in Fig. \ref{fig01}(a) and the corresponding density of states
(DOS)
\begin{equation} \label{Eq24}
Z (E) = \frac{2}{N} \sum\limits_{n \vc{k}} \delta(E_{n} (\vc{k}) - E)
\end{equation}
is displayed in Fig. \ref{fig01}(b). The van Hove peak in the DOS is
broadened by the dispersion of the bandstructure in $z$-direction.

%%%%%%%%%%%%%%%%%%%%%%%%%%%%%%%%%%%%%%%%%%%%%%%%%%%%%%%%%%%%%%%%%%%%%%%%%%%%%%%%%%%%%%%%%
%%%%%%%%%%%%%%%%%%%%%%%%%%%%%%%%%%%%%%%%%%%%%%%%%%%%%%%%%%%%%%%%%%%%%%%%%%%%%%%%%%%%%%%%%
\begin{figure}
\includegraphics[width=\linewidth]{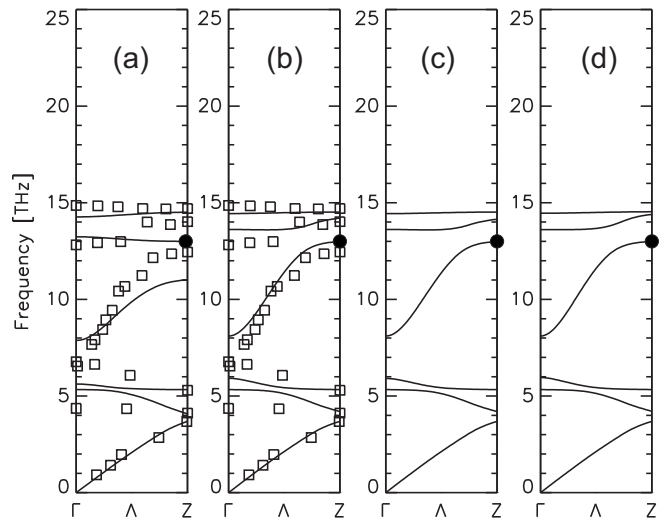}
\caption{Calculated phonon dispersion for LaCuO in the adiabatic
approximation for the $c$-axis polarized $\Lambda_{1}$ modes ($\sim$
(0,0,1)) based on the bandstructure within the 31BM as input for the
proper polarization part $\Pi_{\kappa\kappa'}$ (a). (b)-(d) calculated
$\Lambda_{1}$ modes as in (a) but using the modified 31BM (M31BM) as
discussed in the text as input for $\Pi_{\kappa\kappa'}$. The panels
(b)-(d) represent the phonon dispersion for three different doping
levels $x = 0.156$ ($E_{F} = -0.08$), $x = 0.224$ ($E_{F} = -0.10$) and
$x = 0.297$ ($E_{F} = -0.135$), by applying a rigid-band approximation
of the M31BM to represent the optimally doped to overdoped state of
LaCuO. The experimental results (open squares $\Box$) are taken from
Refs. \onlinecite{Ref31,Ref32}. The full dot ($\bullet$) denotes the
O$_z^Z$ mode. $E_F$ is in units of $eV$.}\label{fig02}
\end{figure}
%%%%%%%%%%%%%%%%%%%%%%%%%%%%%%%%%%%%%%%%%%%%%%%%%%%%%%%%%%%%%%%%%%%%%%%%%%%%%%%%%%%%%%%%%
%%%%%%%%%%%%%%%%%%%%%%%%%%%%%%%%%%%%%%%%%%%%%%%%%%%%%%%%%%%%%%%%%%%%%%%%%%%%%%%%%%%%%%%%%

The corresponding calculated phonon dispersion of LaCuO in the
adiabatic approximation for the $\Lambda_{1}$ modes polarized along the
$c$-axis being most sensitive with respect to the interlayer coupling
is presented in Fig. \ref{fig02}(a). The open squares indicate the
present ''interpretation'' in the literature of the experimental
results \cite{Ref31,Ref32}. This interpretation will be discussed in
detail in Sec. III B. The characteristic experimental features of the
dispersion are the step-like structure of the second highest branch and
most significant the third highest branch with the steep dispersion
from the $\Gamma$-point towards the $Z$-point. Both features are not
well reflected in the calculation based on the typical DFT-LDA-like
bandstructure of the 31BM which underlies the computation of the static
$(\omega = 0)$ proper polarization part $\Pi_{\kappa\kappa'}$ in Eq.
\eqref{Eq18}. As we see in a moment, the reason for this is an
overestimation of the residual $k_{z}$ dispersion of the bands, i.e.
the DFT-LDA-like bandstructure is not anisotropic enough. It should be
remarked that an earlier calculation within the 31BM\cite{Ref46}
neglecting dipole fluctuations as additional electronic polarization
processes besides the charge fluctuations still further enhances the
deviations between the calculated and the measured result for the
$\Lambda_{1}$-modes. Anisotropic DF's being particular important along
the $c$-axis and for ions in the ionic layers \cite{Ref43} have been
taken into account in the calculations shown in Fig. \ref{fig02}(a) and
in all other computations of the phonon dispersion presented in this
paper.

In order to improve the LDA-like electronic BS of the 31BM which
overestimates the $c$-axis coupling we have investigated the effect on
the $\Lambda_{1}$-modes of a modification of the tight-binding
parameters of the 31BM. It turns out that a reduction of the first
neighbour O$_{xy}$-La parameters by 1/6 and of the first neighbour
La-La parameters by 1/3 leads to a much better result for the
calculated $\Lambda_{1}$-modes in adiabatic approximation, see Fig.
2(b-d). The latter results have been obtained for three different
doping levels, where the effect of alloying was treated in rigid-band
approximation by lowering the Fermi level appropriately to accommodate
$x$ holes per primitve cell.

%%%%%%%%%%%%%%%%%%%%%%%%%%%%%%%%%%%%%%%%%%%%%%%%%%%%%%%%%%%%%%%%%%%%%%%%%%%%%%%%%%%%%%%%%
%%%%%%%%%%%%%%%%%%%%%%%%%%%%%%%%%%%%%%%%%%%%%%%%%%%%%%%%%%%%%%%%%%%%%%%%%%%%%%%%%%%%%%%%%
\begin{figure}
\includegraphics[width=\linewidth]{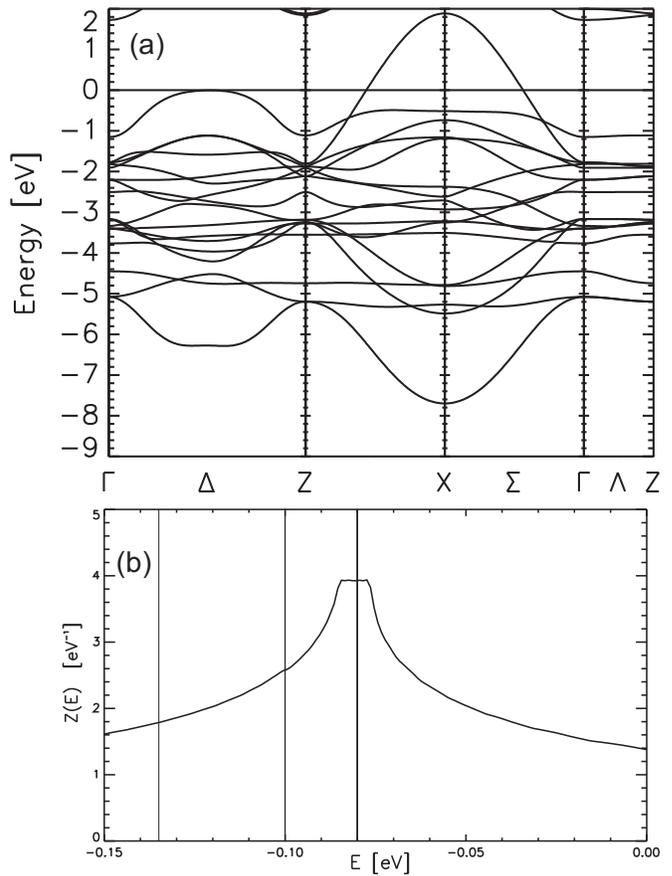}
\caption{Electronic bandstructure $E_{n}(\vc{k})$ of LaCuO in the
modified 31 band model (M31BM) taking into account the enhanced
anisotropy of the real material as compared with the 31BM typical for
LDA (a). Density of states $Z(E)$ in the M31BM (b). The vertical lines
indicate the different doping levels from optimal to overdoping as in
Fig. \ref{fig02}.}\label{fig03}
\end{figure}
%%%%%%%%%%%%%%%%%%%%%%%%%%%%%%%%%%%%%%%%%%%%%%%%%%%%%%%%%%%%%%%%%%%%%%%%%%%%%%%%%%%%%%%%%
%%%%%%%%%%%%%%%%%%%%%%%%%%%%%%%%%%%%%%%%%%%%%%%%%%%%%%%%%%%%%%%%%%%%%%%%%%%%%%%%%%%%%%%%%

The modified electronic BS of LaCuO and the corresponding DOS is shown
in Fig. \ref{fig03}(a) and Fig. \ref{fig03}(b), respectively. The
calculated phonon dispersion is now in good agreement with the
experiment. In particular the characteristic features, i.e. the
$\Lambda_{1}$-branch with the steep dispersion and the step-like
behaviour are well described. Moreover, we find as a consequence of the
enhanced anisotropy a rearrangement of the three $Z$-point modes with
the highest frequencies. While in the 31BM the anomalous O$_z^Z$ mode
is the second highest mode in the more anisotropic modified 31BM
(M31BM) with a weaker coupling along the $c$-axis O$_z^Z$ is the lowest
of the three modes and the end point of the steep branch.

From a comparison of the BS of the 31BM in Fig. \ref{fig01}(a) with the
BS of the M31BM in Fig 3(a) we extract that in the latter case the
saddle point region around $\Delta/2$ is more extended and the width of
the BS is reduced by about 0.5 eV in the energy range shown in the
figure. Important for the charge response along the $c$-axis is the
decrease of the $k_{z}$-dispersion of the bands along the $\Lambda$
direction, which obviously is consistent with the experimental phonon
dispersion of the $\Lambda_{1}$ modes being most sensitive to the
charge dynamics along the $c$-axis. Comparing the calculated results
for the DOS in both models, see Fig. \ref{fig01}(b) and Fig.
\ref{fig03}(b), respectively, we find that the broadening of the van
Hove peak is decreased in the M31BM by about 50 $\%$ because of the
reduced $k_{z}$-dispersion. Simultaneously, the DOS $Z (E)$ around the
peak is enhanced by the increased anisotropy.

Another important effect of the weakening of the interlayer coupling in
our modeling as compared to a typical DFT-LDA bandstructure, leading to
a significant enhancement of the DOS $Z(E_{F})$ at the Fermi energy
$E_{F}$, is the amplification of the proper polarization part
$\Pi_{\kappa\kappa'} (\vc{q})$ on his part. This can be seen in the
long wavelength limit where the sum rule
\begin{equation} \label{Eq25}
\lim\limits_{\vc{q} \to \vec{0}} \sum\limits_{\kappa\kappa'} \Pi_{\kappa\kappa'} (\vc{q}) = Z(E_{F})
\end{equation}
rigorously holds for a metal\cite{Ref39}. Altogether, the results
demonstrate the importance of a correct interlayer-coupling also for
the electronic properties in the CuO layer.

The vertical lines in Fig. \ref{fig03}(b) indicate three doping levels
$x$ in rigid-band approximation. The specific dopings have been
selected to model the optimally doped state of LaCuO with the Fermi
energy $E_{F}$ at the van Hove peak (model OP: $E_{F} = -0.08$ eV, $x =
0.156$) and two overdoped states (model OD1: $E_{F} = -0.10$ eV, $x =
0.224$; model OD2: $E_{F} = -0.135$ eV, $x = 0.297$).

\begin{figure}
\includegraphics[width=\linewidth]{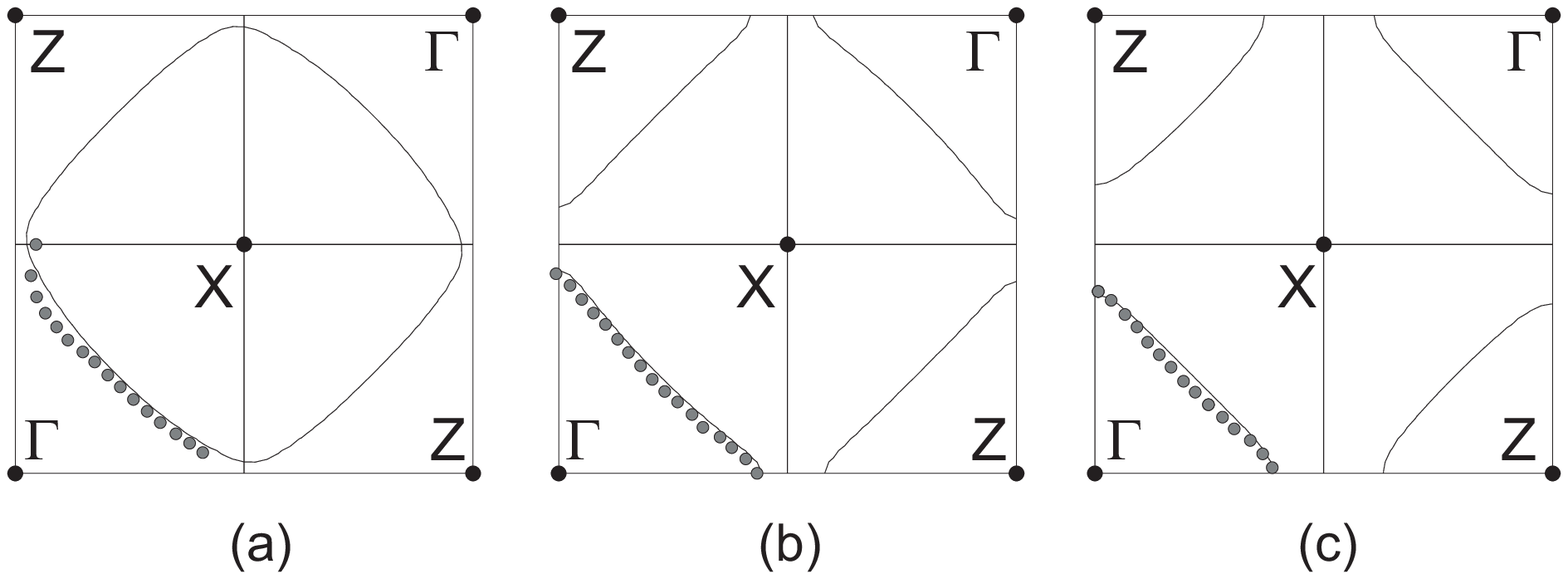}
\caption{Comparison of the doping evolution of the measured Fermi
surface of La$_{2-x}$Sr$_{x}$ CuO$_{4}$ from Ref.
\onlinecite{Ref47,Ref48} for optimal doping ($x = 0.15$) to overdoping
($x = 0.22$; $x = 0.30$) with the corresponding calculated Fermi
surface within the rigid-band approximation of the M31BM, respectively.
The dots indicate the experimental results.}\label{fig04}
\end{figure}

The doping levels chosen in this way allow for a comparison of measured
Fermi surfaces (FS) at these levels \cite{Ref47,Ref48} with the
calculated FS within the M31BM, see Fig. \ref{fig04}. From the
comparison we conclude that the overall features of the BS calculation
in the M31BM are in good agreement with the results obtained by
angle-resolved photoemission spectroscopy (ARPES) over the doping range
considered. Such an agreement obtained within the M31BM and the
consistence with the $\Lambda_{1}$ phonons speaks in favour of this
model with renormalized $c$-axis hopping parameters leading to an
enhanced anisotropy. Moreover, it demonstates the importance of an
experimental verification of a computed electronic bandstructure.
Clearly visible in the experiments and the calculations is the change
of the FS topology and of the related nesting structures upon doping
when passing from the model OP to model OD1 or OD2, respectively.

%%%%%%%%%%%%%%%%%%%%%%%%%%%%%%%%%%%%%%%%%%%%%%%%%%%%%%%%%%%%%%%%%%%%%%%%%%%%%%%%%%%%%%%%%
%%%%%%%%%%%%%%%%%%%%%%%%%%%%%%%%%%%%%%%%%%%%%%%%%%%%%%%%%%%%%%%%%%%%%%%%%%%%%%%%%%%%%%%%%
%%%%%%%%%%%%%%%%%%%%%%%%%%%%%%%%%%%%%%%%%%%%%%%%%%%%%%%%%%%%%%%%%%%%%%%%%%%%%%%%%%%%%%%%%
%%%%%%%%%%%%%%%%%%%%%%%%%%%%%%%%%%%%%%%%%%%%%%%%%%%%%%%%%%%%%%%%%%%%%%%%%%%%%%%%%%%%%%%%%

Strong nesting of the FS can bring about either a charge density wave
or a spin density wave, or possibly both. The tendency toward Fermi
surface driven instabilities frequently is anounced by maxima of the so
called (noninteracting) suszeptibility
\begin{equation} \label{Eq26}
\Pi_{0} (\vc{q}) = -\frac{2}{N} \sum\limits_{n,n' \atop \vc{k}}
\frac{f_{n'} (\vc{k} + \vc{q}) - f_{n} (\vc{k})}{E_{n'} (\vc{k} +
\vc{q}) - E_{n} (\vc{k})} ,
\end{equation}
which can be obtained form Eq. \eqref{Eq18} by equating all the
expansion coefficients $C_{\kappa n} (\vc{k})$ to one. It is obvious
from this form that possible FS-nesting at certain $\vc{q}$ vectors is
reflected as a maximum of this function.

From inspection of the FS in Fig. \ref{fig04} we can expect that
approximate nesting vectors around the $X$-point of the BZ will lead to
corresponding maxima in $\Pi_{0} (\vc{q})$. For the optimally case
(model OP) we find a maximum at the $X$-point and in the  overdoped
state (model OD1 and OD2) the maximum is progressively shifted toward
the $\Gamma$ point. The position of the maximum for model OP is
significant because it coincides with the ordering wave vector in the
antiferromagnetic state. For the overdoped state, the ordering wave
vector (magnetic scattering peak) is predicted from our calculations
for $\Pi_{0}$ to move inward from the $X$-point. Quite general, in the
doped cuprates the spin  fluctuations (SF's) are antiferromagnetic in
origin and arise from both, nesting properties of the FS
(spin-density-wave type SF's) and from correlation driven
nearest-neighbour antiferromagnetic superexchange because of the
proximity of a long-range ordered antiferromagnetic state in the
undoped parent compounds.

%%%%%%%%%%%%%%%%%%%%%%%%%%%%%%%%%%%%%%%%%%%%%%%%%%%%%%%%%%%%%%%%%%%%%%%%%%%%%%%%%%%%%%%%%
%%%%%%%%%%%%%%%%%%%%%%%%%%%%%%%%%%%%%%%%%%%%%%%%%%%%%%%%%%%%%%%%%%%%%%%%%%%%%%%%%%%%%%%%%
\begin{figure*}
\includegraphics[width=\linewidth]{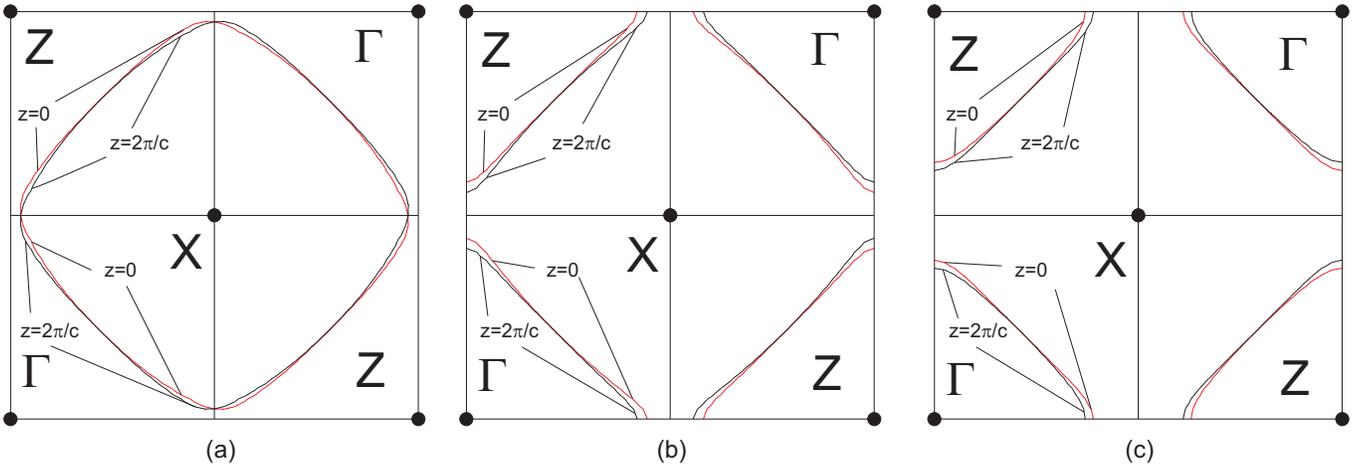}
\caption{Fermi surface maps for different $k_{z}$ values ($k_{z} \equiv
z = 0$ and $k_{z} \equiv z = \frac{2 \pi}{c}$) as a function of
$\vc{k}_{\|} = (k_{x}, k_{y})$ for different doping levels in
rigid-band approximation of the M31BM. Model OP ($x = 0.156$) (a),
model OD1 ($x = 0.224$) (b) and model OD2 ($x = 0.297$)
(c).}\label{fig05}
\end{figure*}
%%%%%%%%%%%%%%%%%%%%%%%%%%%%%%%%%%%%%%%%%%%%%%%%%%%%%%%%%%%%%%%%%%%%%%%%%%%%%%%%%%%%%%%%%
%%%%%%%%%%%%%%%%%%%%%%%%%%%%%%%%%%%%%%%%%%%%%%%%%%%%%%%%%%%%%%%%%%%%%%%%%%%%%%%%%%%%%%%%%

In Fig. \ref{fig05} we display for the models OP, OD1 and OD2 FS maps
for different $k_{z}$ values ($k_{z} = 0, k_{z} = \frac{2\pi}{c})$ as a
function of $\vc{k} = (k_{x}, k_{y})$. This allows for example to map
the $k_{x}$-, $k_{y}$-dependence of the $c$-axis dispersion. The
calculations demonstrate that the effect of the $k_{z}$ dispersion
vanishes along the nodal direction and that it increases as one moves
towards the antinodal region. The three-dimensionality associated with
the $k_{z}$-dispersion, usually neglected in discussing the physics of
the cuprates, has been shown to play a key role in shaping the ARPES
spectra \cite{Ref36,Ref49}, because the residual $k_{z}$-dispersion of
bands in a quasi-two-dimensional material will induce an irreducible
linewidth in ARPES peaks. On the other hand detecting such $k_{z}$
related linewidth in the ARPES spectra as found for LaCuO \cite{Ref36}
establishes the existence of coherent $c$-axis transport.

Finally to get also a more global impression of the magnitude of the
enhanced anisotropy in the M31BM as compared with a typical DFT-LDA
based first principles result we compare some FS parameters being
important for transport properties like the Drude plasma energy tensor
and the Fermi velocity tensor. The Drude tensor is defined as
\begin{equation} \label{Eq27}
\hbar \Omega_{p, ij} = \left(\frac{8 \pi}{N V_{z}} \sum\limits_{\vc{k}
n} \delta (E_{n}(\vc{k}) - E_{F}) v_{\vc{k} n,i} v_{\vc{k} n,
j}\right)^{\frac{1}{2}} ,
\end{equation}
and the Fermi velocity tensor ist given by
\begin{equation} \label{Eq28}
\left< v_{F, ij}^{2}\right>^{\frac{1}{2}} = \left(\frac{2}{N}
\sum\limits_{\vc{k} n} \Theta (E_{n}(\vc{k}) - E_{F}) v_{\vc{k} n, i}
v_{\vc{k} n, j}\right)^{\frac{1}{2}},
\end{equation}
with
\begin{equation} \label{Eq29}
\vc{v}_{\vc{k} n} = \frac{1}{\hbar} \frac{\partial
E_{n}(\vc{k})}{\partial \vc{k}} .
\end{equation}
The outcome from our model OP for the optimally doped state is compared
in Table \ref{tab1} with the corresponding results calculated for
optimally doped LaCuO within LAPW \cite{Ref50,Ref51}. From the table we
extract an enhancement of the anisotropy ratio $A_{\Omega} =
\Omega_{p,xx}/\Omega_{p,zz}$ and of $A_{v_{F}} =
(v^{2}_{{F,xx}})^{1/2}/(v^{2}_{F,zz})^{1/2}$ by about a factor 5 in the
M31BM as compared to the LAPW calculation.

\subsection{Nonadiabation phonon calculations and phonon-plasmon coupling}
\begin{figure}
\includegraphics[width=\linewidth]{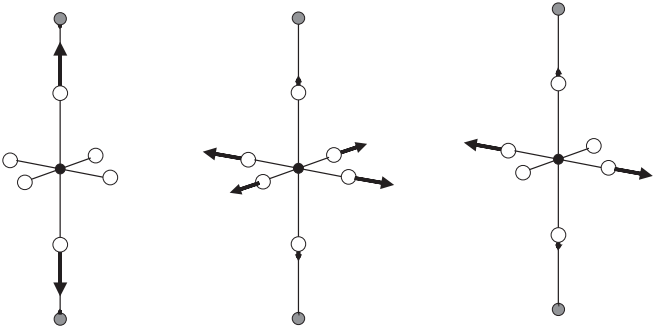}
\caption{Displacement patterns of the high-frequency oxygen
bond-stretching modes (OBSM) of LaCuO. Left: apex-oxygen breathing mode
(O$_z^Z$), middle: planar breathing mode ($O^{X}_{B}$), right:
half-breathing mode ($\Delta_{1}/2$).}\label{fig06}
\end{figure}

We prelude this section with a discussion of the anomalous behaviour of
the apex-oxygen bond-stretching mode in LaCuO at the $Z$-point of the
BZ (O$_z^Z$) polarized perpendicular to the CuO plane, see Fig.
\ref{fig06}. From the experimental side, it took quite some time until
this mode could be assessed \cite{Ref32} because of a massive
line-broadening of about 4 THz in the metallic phase at optimal doping
and its very large softening of about 5.5 THz when passing from the
insulating to the metallic state \cite{Ref31,Ref32}. The strong
softening of this mode across the insulator-metal transition was
predicted prior to the experimental observation, see Refs.
\cite{Ref39,Ref42,Ref28}.

\begin{table}
\begin{tabular}{c|cccccc}
   &  $\hbar\Omega_{\text{p},xx}$ & $\hbar\Omega_{\text{p},zz}$ & $\left<v^2_{\text{F},xx}\right>^{1/2}$ & $\left<v^2_{\text{F},zz}\right>^{1/2}$ & $A_\Omega$ & $A_{\text{v}_\text{F}}$  \\\hline
LAPW  & 701.21 & 132.99  & 2.20  &  0.41  &  5.27  &   5.37 \\
OP    & 648.60 & 25.25  & 2.99 &  0.11 &  25.69  & 27.97
\end{tabular}
\caption{Comparison of the Fermi surface parameters (Drude plasma
energy tensor, Fermi velocity) and the anisotropy ratios $A_{\Omega} =
\Omega_{p,xx}/\Omega_{p,zz}$; $A_{v_{F}} = \langle
v^{2}_{F,xx}\rangle^{1/2}/\langle v^{2}_{F,zz}\rangle^{1/2}$ for
optimally doped LaCuO between a calculation in LAPW \cite{Ref50,Ref51}
and the M31BM (model OP), respectively. $\Omega_{p,ij}$ is given in
units of THz and the Fermi velocity $\langle v_{F,ij}^{2}\rangle^{1/2}$
in units of $10^{7}$ cm/s.}\label{tab1}
\end{table}

%%%%%%%%%%%%%%%%%%%%%%%%%%%%%%%%%%%%%%%%%%%%%%%%%%%%%%%%%%%%%%%%%%%%%%%%%%%%%%%%%%%%%%%%%
%%%%%%%%%%%%%%%%%%%%%%%%%%%%%%%%%%%%%%%%%%%%%%%%%%%%%%%%%%%%%%%%%%%%%%%%%%%%%%%%%%%%%%%%%
%%%%%%%%%%%%%%%%%%%%%%%%%%%%%%%%%%%%%%%%%%%%%%%%%%%%%%%%%%%%%%%%%%%%%%%%%%%%%%%%%%%%%%%%%
%%%%%%%%%%%%%%%%%%%%%%%%%%%%%%%%%%%%%%%%%%%%%%%%%%%%%%%%%%%%%%%%%%%%%%%%%%%%%%%%%%%%%%%%%

Optical modes like O$_z^Z$ are of special importance for the cuprates
because of their strong \textit{nonlocal} coupling to the electrons due
to the long-ranged Coulomb interaction poorly screened along the
$c$-axis in these compounds, even in the metallic state. As a
consequence nonlocal, nonadiabatic polar electron-phonon coupling
effects become substantial. These high-frequency $c$-axis polarized
optical phonons are basically unscreened. Quite recently such a
long-ranged polar electron-phonon interaction (EPI) has been identified
as very essential also for pairing in the cuprate superconductors
\cite{Ref52,Ref53}.

Displacements of the ions in the ionic layers of the cuprates, in
particular those of the O$_z^Z$ mode, bring about large changes of the
potential felt by the electrons (holes) in the CuO plane which on her
part are responsible for the superconductivity and the unusual normal
state effects. So, the $c$-axis polarized optical phonons become
important for the renormalization of the electrons in the CuO plane,
besides the renormalization due to strong correlation effects. Thus,
both, the short-ranged and the long-ranged part of the Coulomb
interaction is significant for the physics in the HTSCs.

The manifestation of the coupling between electrons and polar $c$-axis
phonons in the self-energy of the nodal quasiparticles in Bi2201 has
been pointed out recently in Refs. \onlinecite{Ref17,Ref54}. Note, that
the long-ranged Coulomb coupling perpendicular to the CuO plane is very
special for the HTSC's and would not be possible in a conventional
three-dimensional metal or superconductor because of local screening by
a high-density electron gas. As subsequently discussed this long-ranged
polar interaction does not only probe the in-plane electron properties
but also leads to a strong coupling between electrons and $c$-axis
polarized  phonon modes.

In the insulating parent compound O$_z^Z$ is found at about 17 THz
while in the optimally doped material the experimental value is
estimated at about 11.5 THz \cite{Ref31}. However, the large linewidth
of about 3-4 THz \cite{Ref31,Ref32} and the presence of another
$Z$-point mode at about 14 THz makes it difficult to pin down the exact
position of O$_z^Z$ in the metallic phase.

In the following we will demonstrate within the M31BM that due to the
nonlocal, nonadiabatic charge response two coupled phonon-plasmon modes
of O$_z^Z$-type arise at 9.09 THz and 13.86 THz, respectively, and that
we can attribute to O$_z^Z$ an {\em adiabatic} frequency of 12.96 THz.
Before discussing these results in more detail we present in retrospect
a short explanation of the large softening of O$_z^Z$ during the
insulator-metal transition, for details see e.g. Ref.
\onlinecite{Ref28} and earlier references therein.

In the insulating state O$_z^Z$ can be shown to induce only an {\em
intralayer} charge transfer between the copper and oxygen orbitals such
that local charge neutrality of the cell is maintained under a
perturbation due to O$_z^Z$. In this way the charge rearrangement
leading to screening is considerably restricted and thus we find a high
calculated frequency of O$_z^Z$ in the insulating state in agreement
with experiment. Contrarily, in the metallic state no such restriction
is present and O$_z^Z$ generates CF's at Cu and O$_{xy}$ that have the
same  sign in the whole CuO layer. This ultimately leads to an
(instantaneous) interlayer charge transfer in adiabatic approximation
which provides an effective screening mechanism for the long-ranged
Coulomb interaction and produces the anomalous softening of O$_z^Z$
during the insulator-metal transition in the calculations consistent
with the measurements. Eventually, the instantaneous interlayer charge
transfer in the adiabatic approximation is replaced by a collective
dynamic charge transfer within the phonon-plasmon scenario in a small
region around the $c$-axis, including of course the $Z$-point. Such a
scenario will be investigated next within the M31BM.

%%%%%%%%%%%%%%%%%%%%%%%%%%%%%%%%%%%%%%%%%%%%%%%%%%%%%%%%%%%%%%%%%%%%%%%%%%%%%%%%%%%%%%%%%
%%%%%%%%%%%%%%%%%%%%%%%%%%%%%%%%%%%%%%%%%%%%%%%%%%%%%%%%%%%%%%%%%%%%%%%%%%%%%%%%%%%%%%%%%
\begin{figure}
\includegraphics[width=\linewidth]{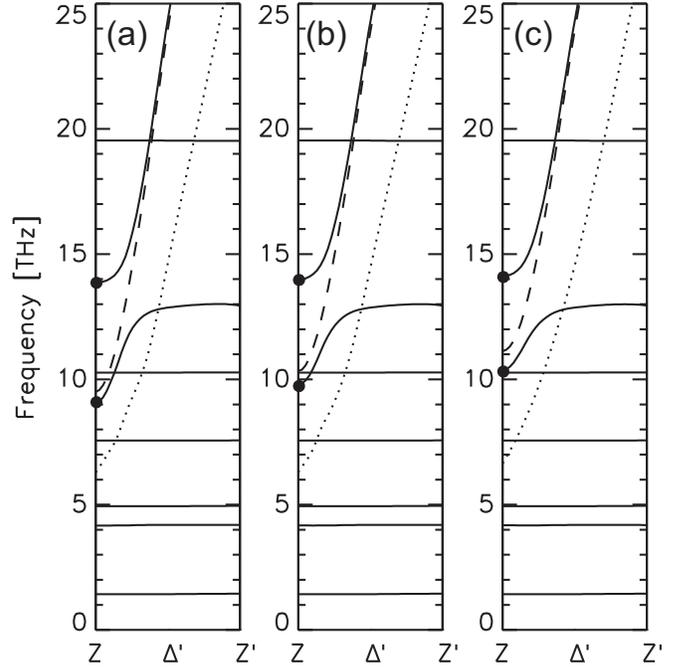}
\caption{Calculated coupled phonon-plasmon dispersion of the
$\Delta'_{1}$, ($O^{Z'}_{z}$) modes ($-\!\!-$) from the $Z$-point
($\varepsilon = 0$) along $\Delta' = \left(\varepsilon \frac{2\pi}{a},
0, \frac{2\pi}{c}\right)$ to $\varepsilon = 0.02$ ($Z'$) for the models
OP (a), OD1 (b) and OD2 (c), respectively. The coupling O$_z^Z$ modes
at $Z$ are represented by $\bullet$; $-\,-$: free plasmon branch;
$\cdots$ borderline for damping due to electron-hole
decay.}\label{fig07}
\end{figure}
%%%%%%%%%%%%%%%%%%%%%%%%%%%%%%%%%%%%%%%%%%%%%%%%%%%%%%%%%%%%%%%%%%%%%%%%%%%%%%%%%%%%%%%%%
%%%%%%%%%%%%%%%%%%%%%%%%%%%%%%%%%%%%%%%%%%%%%%%%%%%%%%%%%%%%%%%%%%%%%%%%%%%%%%%%%%%%%%%%%

\begin{figure}
\begin{minipage}{0.49\linewidth}%
\includegraphics[]{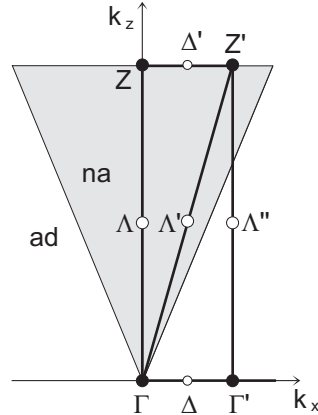}%
\end{minipage}%
\begin{minipage}{0.49\linewidth}%
\caption{Schematic representation of the nonadiabatic region in the
($k_{x}$, $k_{z}$) plane with the directions $\Delta' =
\left(\varepsilon \frac{2\pi}{a}, 0, \frac{2\pi}{c}\right)$, $\Lambda'
= \zeta \left(\varepsilon \frac{2\pi}{a}, 0, \frac{2\pi}{c}\right)$ and
$\Lambda'' = \left(\varepsilon \frac{2\pi}{a}, 0, \zeta
\frac{2\pi}{c}\right)$.
$\zeta \in [0,1]$. na: nonadiabatic region; ad: adiabatic region.}\label{fig08}%
\end{minipage}%
\end{figure}

In Fig. \ref{fig07} we present our results of the dispersion of the
coupled phonon-plasmon modes in the small region around the $c$-axis
characterized by a nonadiabatic charge response. Specifically, we have
calculated the coupled mode dispersion of the $\Delta_{1}'$ modes from
the $Z$-point along $\Delta'$ to the $Z' = \left(\varepsilon
\frac{2\,\pi}{a}, 0, \frac{2\,\pi}{c}\right)$ point with $\varepsilon =
0.02$, see Fig. \ref{fig08}. The mixed mode dispersion in Fig.
\ref{fig07}(a) is for the optimally doped case of LaCuO (model OP),
Fig. \ref{fig07}(b) and 7(c) show the results for the overdoped
material, i.e. for model OD1 and OD2, respectively. The strongly
coupling O$_z^Z$ modes are indicated as full dots. The broken line is
the dispersion of the free-plasmon branch, calculated from Eq.
\eqref{Eq23} and the dotted line is the borderline for damping due to
electron-hole decay, i.e. ${\text{max} \atop \vc{k} \in
\text{BZ}}~(E_{n} (\vc{k}) - E_{n} (\vc{k} + \vc{q}))$ for the band
crossing the Fermi level. We observe a slight increase of the frequency
of the free-plasmon branch upon doping which indicates a small
reduction of the anisotropy.

%%%%%%%%%%%%%%%%%%%%%%%%%%%%%%%%%%%%%%%%%%%%%%%%%%%%%%%%%%%%%%%%%%%%%%%%%%%%%%%%%%%%%%%%%
%%%%%%%%%%%%%%%%%%%%%%%%%%%%%%%%%%%%%%%%%%%%%%%%%%%%%%%%%%%%%%%%%%%%%%%%%%%%%%%%%%%%%%%%%
%%%%%%%%%%%%%%%%%%%%%%%%%%%%%%%%%%%%%%%%%%%%%%%%%%%%%%%%%%%%%%%%%%%%%%%%%%%%%%%%%%%%%%%%%
%%%%%%%%%%%%%%%%%%%%%%%%%%%%%%%%%%%%%%%%%%%%%%%%%%%%%%%%%%%%%%%%%%%%%%%%%%%%%%%%%%%%%%%%%

The phonon-plasmon scenario  is exemplified for optimally doped LaCuO
in Fig. \ref{fig07}(a) corresponding to the model OP for the proper
polarization part. From inspection of the figure we can learn two
important facts. Firstly, the range of the region in $\vc{q}$-space
with a nonadiabatic charge response can be estimated and, secondly, the
massive line broadening of the O$_z^Z$ mode can be understood. The full
dots at the $Z$-point are the phonon-like and plasmon-like O$_z^Z$(na)
mode at 9.09 THz and 13.86 THz, respectively. We denote the lower mode
as phonon-like because this mode connects to the adiabatic
$O^{Z'}_{z}$(ad) phonon in the adiabatic region. $O^{Z'}_{z}$(ad) has
virtually the same frequency as the O$_z^Z$(ad) mode calculated in
adiabatic approximation, i.e. O$_z^Z$(ad) = 12.96 THz. From the figure
we can extract that the region of nonadiabatic charge response,
characterized by the steep phonon-like branch is very small. It can be
estimated at about $\varepsilon \approx 0.01$.

The plasmon-like branch starts at the higher O$_z^Z$(na) mode, rapidly
leaves the frequency range of the phonon spectrum within the
nonadiabatic region and becomes more and more plasmon-like as it
approaches the free-plasmon dispersion.

The experimentally measured massive line-broadening for O$_z^Z$ of
about 3-4 THz \cite{Ref31,Ref32} in the optimally doped probe can be
understood from the calculated phonon-plasmon scenario displayed in
Fig. \ref{fig07}(a). Experimentally, there is a limited wavevector
resolution in the transverse direction perpendicular to the $c$-axis
which is on average $\varepsilon \approx 0.03$ (Ref.
\onlinecite{Ref55}). Thus, the relevant frequency range sampled by the
neutron scattering experiment is over the steep branch and is between
the two nonadiabatic O$_z^Z$(na) modes leading to an estimated
broadening of about 3.9 THz. Because the region with a metallic,
adiabatic charge response outweights by a factor of three the
nonadiabatic region we can attribute to the broad O$_z^Z$ mode an
adiabatic frequency of about 12.9 THz.

\begin{table*}
\begin{tabular}{c|cccccccccc}
$\epsilon$ & $Z$(ad) & $Z$(na) & 0.0001 & 0.001 & 0.002 & 0.003 & 0.004
& 0.007 & 0.01 & 0.02\\\hline\hline $\omega$ & 12.96 & 9.09 & 9.08 &
9.27
& 9.84 & 10.59 & 11.41 & 12.61 & 12.86 & 12.94 \\
O$_\text{z}$(z) & 0.69 & 0.66 & 0.66 & 0.67 & 0.69 & 0.70 & 0.71 & 0.69
& 0.69 & 0.69\\
La(z) & -0.15 & 0.25 & 0.25 & 0.22 & 0.14 & 0.05 & -0.04 & -0.13 &
-0.15 & -0.15\\\hline $\omega$ & 12.96 & 13.86 & 13.86 &
13.90 & 14.02 & 14.28 & 14.81 & 18.69 & 24.38 & $-$\\
O$_\text{z}$(z) & 0.69 & 0.68 & 0.68 & 0.68 & 0.68 & 0.67 & 0.67 & 0.61
& 0.59  & $-$\\
La(z) & -0.15 & -0.20 & -0.20 & -0.20 & -0.21 & -0.22 & -0.24 & -0.33 &
-0.38 & $-$
\end{tabular}
\caption{Calculated displacement amplitudes of the phonon-like and
plasmon-like $O^{Z'}_{z}$ modes in model OP along the $\Delta' =
\left(\varepsilon \frac{2\pi}{a}, 0, \frac{2\pi}{c}\right)$ direction
between $\varepsilon = 0$ ($Z$-point) and $\varepsilon = 0.02$. The
frequency is given in units of THz. The left most data set is the
result for O$_z^Z$ in the adiabatic limit. The plasmon-like mode for
$\varepsilon = 0.02$ has not been calculated because of its frequency
is far out of the relevant frequency range, see Fig.
\ref{fig07}(a).}\label{tab2}
\end{table*}

The calculated character of the coupled modes along $\Delta'$ can be
obtained from Table \ref{tab2} where the displacement amplitudes for
the phonon-like and plasmon-like mode is given in terms of relative
displacements of the apex oxygen $O_{z}$ and the La ion along the
$c$-axis. At the $Z$-point $O_{z}$ and La are vibrating in phase for
the phonon-like mode and out of phase for the plasmon-like mode. In
case of the phonon-like mode the amplitude of O$_z^Z$ remains more or
less constant while the amplitude for La decreases with increasing
$\varepsilon$ and finally runs out of phase with $O_{z}$ in the
adiabatic region. In the plasmon-like O$_z^Z$ mode the $O_{z}$ and La
ions always vibrate out of phase. Increasing $\varepsilon$ the
amplitude of O$_z^Z$ decreases while the magnitude of the amplitude of
La increases. The different phase relation of the two modes is
reflected in the corresponding induced charge redistributions as
discussed below.

The size $\varepsilon \approx 0.01$ of the nonadiabatic region
predicted  by our calculation within model OP is too small for an
experimental observation of the dispersion of the coupled
phonon-plasmon modes shown in Fig. \ref{fig07}. Only sampling the
nonadiabatic region on average as mentioned above seems possible at
present.

Inspection of Fig. \ref{fig07}(b) and Fig. \ref{fig07}(c) for the
overdoped states leads to about the same size of the nonadiabatic
sector as in the optimally doped case and to the same adiabatic
frequency for $O^{Z'}_{z}$(ad) and O$_z^Z$(ad), respectively. However,
the broadening of O$_z^Z$ is predicted by the calculation to decrease
as compared to the optimally doped case. A decrease of the linewidth of
O$_z^Z$ with practically no change of the frequency has also been found
quite recently for overdoped LaCuO in the experiments \cite{Ref56}.
Thus, our prediction within the rigid-band approximation is conform
with the measurements. From Fig. \ref{fig07}(b) and Fig. \ref{fig07}(c)
the width of O$_z^Z$ is 3.23 THz for $x \approx 0.22$ and 2.66 THz for
$x \approx 0.3$, respectively. Last but not least we can read off from
Fig. \ref{fig07} a doping dependence of the phonon-like O$_z^Z$ mode in
which the mode energy decreases with reduced doping (OD2: 10.31 THz;
OD1: 9.74 THz; OP: 9.09 THz).

\begin{figure*}
\includegraphics[width=\linewidth]{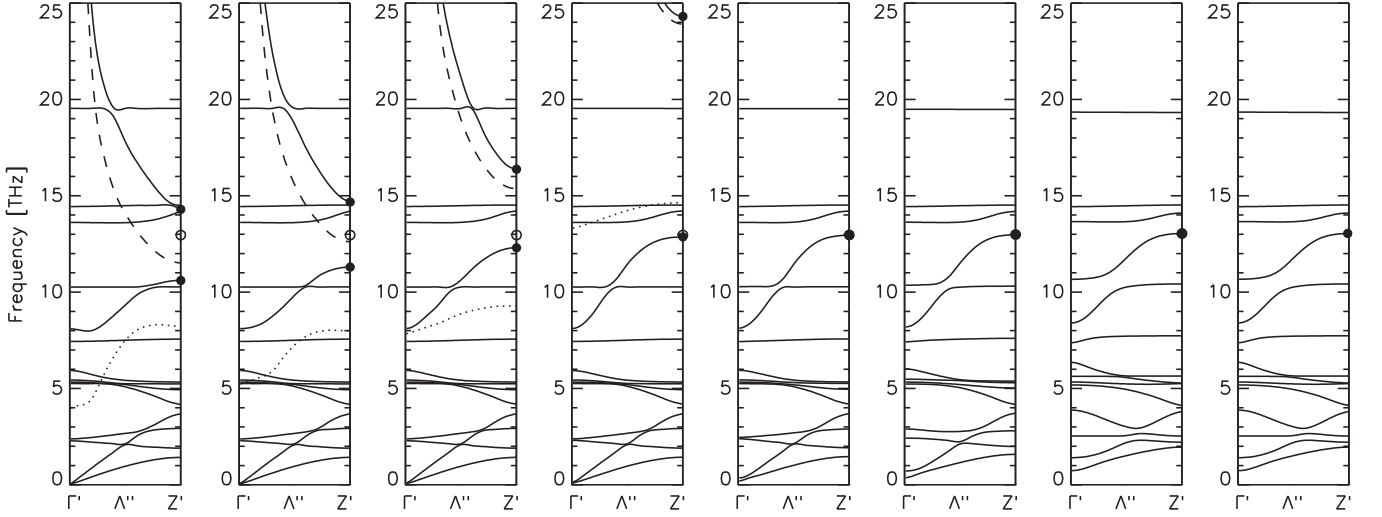}
\caption{Nonadiabatic coupled phonon-plasmon dispersion as calculated
with model OP for the proper polarization part, representing the
optimally doped state of LaCuO from $\Gamma' = \left(\varepsilon
\frac{2\pi}{a}, 0, 0\right)$ to $Z'$ along the $\Lambda'' = \left(
\varepsilon \frac{2\pi}{a}, 0, \zeta \frac{2\pi}{c}\right)$ direction.
In the different panels of the figure from left to right the following
values of $\varepsilon$ have been used: $\varepsilon = 0.003, 0.004,
0.006, 0.01, 0.025, 0.05, 0.100$. The right most panel shows the
results of a calculation in adiabatic approximation for $\varepsilon =
0.100$. Only the $\Lambda''$ branches ($-\!\!-$) coupling to the charge
fluctuations are shown. $-\,-$: free-plasmon branch; $\cdots$:
borderline for damping. Dots at $Z'$: $\bullet$ $O^{Z'}_{z}$(na);
$\circ$ $O^{Z'}_{z}$(ad). na: nonadiabatic; ad:
adiabatic.}\label{fig09}
\end{figure*}

Now we comment on the present interpretation of the measured dispersion
of the steep $\Lambda_{1}$-branch displayed in Fig. \ref{fig02}. We
just have shown through our calculation of the phonon-plasmon
dispersion along the $\Delta'$ direction that the nonadiabatic region
around the $c$-axis is so small that it cannot be resolved by the INS
experiments and only an average of the dispersion can be measured. So,
we investigate in detail the coupled mode dispersion along the two
directions $\Lambda'' = \left(\varepsilon \frac{2\pi}{a}, 0, \zeta
\frac{2\pi}{c}\right)$ in Fig. \ref{fig09} and $\Lambda' = \zeta
\left(\varepsilon \frac{2\,\pi}{a}, 0, \frac{2\pi}{c}\right)$ in Fig.
10 nearby the $c$-axis in order to study the transition from the region
of nonadiabatic to adiabatic charge response together with the change
of the mode behavior during this transition.

%%%%%%%%%%%%%%%%%%%%%%%%%%%%%%%%%%%%%%%%%%%%%%%%%%%%%%%%%%%%%%%%%%%%%%%%%%%%%%%%%%%%%%%%%
%%%%%%%%%%%%%%%%%%%%%%%%%%%%%%%%%%%%%%%%%%%%%%%%%%%%%%%%%%%%%%%%%%%%%%%%%%%%%%%%%%%%%%%%%
%%%%%%%%%%%%%%%%%%%%%%%%%%%%%%%%%%%%%%%%%%%%%%%%%%%%%%%%%%%%%%%%%%%%%%%%%%%%%%%%%%%%%%%%%
%%%%%%%%%%%%%%%%%%%%%%%%%%%%%%%%%%%%%%%%%%%%%%%%%%%%%%%%%%%%%%%%%%%%%%%%%%%%%%%%%%%%%%%%%

The frequency of the plasmon near $\Gamma'$ in Fig. \ref{fig09} is
always very high and far outside the range of the phonon spectrum.
Consequently the phonon dynamics is adiabatic in this $\vc{q}$-space
region. The phonon-like nonadiabatic $O^{Z'}_{z}$ mode (lower full dot)
converges to its adiabatic value (open circle) at about $\varepsilon =
0.01$, while the plasmon-like $O^{Z'}_{z}$ mode (upper full dot)
rapidly leaves the range of the spectrum. Simultaneously the dynamics
becomes adiabatic. Compare with the calculation of the phonon
dispersion in the right most panel where the static ($\omega = 0$)
approximation for $\Pi_{\kappa\kappa'}$ has been used. Likewise as in
the investigations along the $\Delta'$ direction the nonadiabatic
region can be characterized by $\varepsilon \approx 0.01.$ and the
present resolution limit for INS does not allow to resolve the
mixed-phonon plasmon dispersion. Only the result of averaging
$\Lambda''$ between $\Gamma'$ and $Z'$ can be detected. In such a
procedure the small nonadiabatic part close to the $c$-axis will be
outweighted in the measurement by a significant larger part where the
dispersion is nearly adiabatic. In particular we extract the
development of the dispersion which leads to the steep $\Lambda_{1}$
branch along $\Lambda$ found in our calculations in the adiabatic
limit. From this discussion it becomes clear how the measured data
points in Fig. \ref{fig02} should be interpreted, namely as an average
over the coupled mode dispersion with a dominant adiabatic
contribution.

\begin{figure*}
\includegraphics[width=\linewidth]{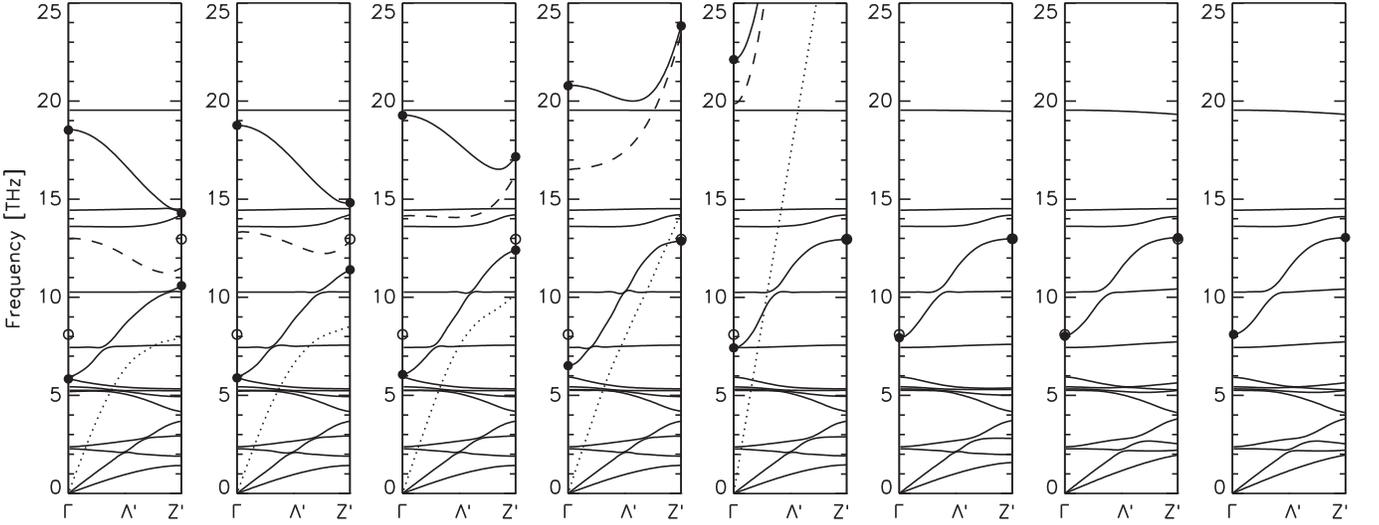}
\caption{Nonadiabatic coupled phonon-plasmon dispersion as calculated
with model OP for the optimally doped state of LaCuO from $\Gamma$ to
$Z' = \left(\varepsilon \frac{2\pi}{a}, 0, \frac{2\pi}{c}\right)$ along
the $\Lambda' = \zeta \left(\varepsilon \frac{2\pi}{a}, 0,
\frac{2\pi}{c}\right)$ direction. In the different panels from left to
right the following values for $\varepsilon$ have been used:
$\varepsilon = 0.003, 0.004, 0.006, 0.010, 0.025, 0.050, 0.100$. The
right most panel shows the result of a calculation within adiabatic
approximation for $\varepsilon = 0.100$. Only the $\Lambda'$ branches
($-\!\!\!-$) coupling to the charge fluctuations are shown. $-\,-$:
free plasmon branch; $\cdots$: borderline for damping. Dots at $Z'$:
$\bullet$ $O^{Z'}_{z}$(na), $\circ$ $O^{Z'}_{z}$(ad). Dots at $\Gamma$:
$\bullet$ A$^{\text{LO}}_{2u}$ (ferro, na); $\circ$
A$^{\text{TO}}_{2u}$ (ferro, ad). na: nonadiabatic; ad:
adiabatic.}\label{fig10}
\end{figure*}

Concerning the steep $\Lambda_{1}$ branch the same conclusion can be
drawn from the inspection of the calculated results along the
$\Lambda'$ direction displayed in Fig. \ref{fig10}. Again the
phonon-like $O^{Z'}_{z}$ mode matches the adiabatic result at about
$\varepsilon \approx 0.01$ and the dispersion with larger $\varepsilon$
values is virtually adiabatic. At the $\Gamma$-point two coupling
longitudinal $A_{2u}$ modes (full dots) occur which we denote as
ferroelectric modes (FM) because of their intrinsic displacement
patterns where the oxygen anions are vibrating coherently against the
cations in the lattice. As a consequence the electric dipole moments
generated by the motion add constructively to a large value. So, we can
expect a large oscillator strength and as seen in the experiments
\cite{Ref14} the FM dominates the infrared response for polarization
along the $c$-axis not only in the insulating state of LaCuO but also
in the well doped metallic state. As already mentioned, such an optical
activity in the \textit{metallic} phase cannot be explained using the
adiabatic approximation for the calculation of the phonon dispersion as
is usually done by applying static DFT for the metal, because there
will be no LO-TO splitting (A$_{2u}$ splitting in the present case)
being a measure of the oscillator strength in such a calculation.

%%%%%%%%%%%%%%%%%%%%%%%%%%%%%%%%%%%%%%%%%%%%%%%%%%%%%%%%%%%%%%%%%%%%%%%%%%%%%%%%%%%%%%%%%
%%%%%%%%%%%%%%%%%%%%%%%%%%%%%%%%%%%%%%%%%%%%%%%%%%%%%%%%%%%%%%%%%%%%%%%%%%%%%%%%%%%%%%%%%
%%%%%%%%%%%%%%%%%%%%%%%%%%%%%%%%%%%%%%%%%%%%%%%%%%%%%%%%%%%%%%%%%%%%%%%%%%%%%%%%%%%%%%%%%
%%%%%%%%%%%%%%%%%%%%%%%%%%%%%%%%%%%%%%%%%%%%%%%%%%%%%%%%%%%%%%%%%%%%%%%%%%%%%%%%%%%%%%%%%

On the other hand, the observed infrared response in the metallic phase
can be understood from our nonadiabatic results in Fig. \ref{fig10}.
Here we find a very large A$_{2u}$ splitting for the longitudinal
plasmon-like FM (upper full dot at $\Gamma$) at 18.41 THz and the
corresponding transverse FM (open circle at $\Gamma$) with a calculated
frequency of 8.1 THz and 7.4 THz in the experiments \cite{Ref43},
respectively. Simultaneously, there is a smaller splitting between the
latter mode and the longitudinal phonon-like FM (lower full dot at
$\Gamma$) at 5.80 THz. With increasing $\varepsilon$, i.e. when
approaching the region of adiabatic charge response, the LO-TO
splitting is closed from {\em below} and the plasmon-like branch with
the plasmon-like FM at $\Gamma$ and the plasmon-like $O^{Z'}_{z}$ mode
at $Z'$ rapidly disappears out of the phonon spectrum. Moreover, the
phonon dispersion becomes adiabatic and the signature of the steep
$\Lambda_{1}$ branch appears in the $\Lambda'$ direction.

Summarizing, it should be emphasized that the presence of the large
nonlocal long-ranged polar nonadiabatic electron-phonon coupling which
leads to the phonon-plasmon scenario is also reflected in the measured
infrared response.

%%%%%%%%%%%%%%%%%%%%%%%%%%%%%%%%%%%%%%%%%%%%%%%%%%%%%%%%%%%%%%%%%%%%%%%%%%%%%%%%%%%%%%%%%
%%%%%%%%%%%%%%%%%%%%%%%%%%%%%%%%%%%%%%%%%%%%%%%%%%%%%%%%%%%%%%%%%%%%%%%%%%%%%%%%%%%%%%%%%
\begin{figure}
\begin{minipage}{0.3\linewidth}%
\includegraphics[width=\linewidth]{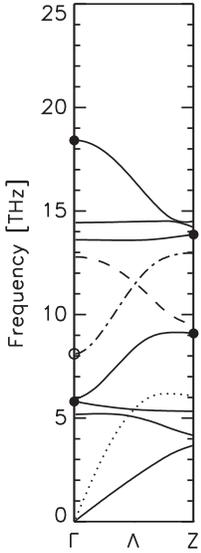}%
\end{minipage}\hspace{0.1\linewidth}%
\begin{minipage}{0.49\linewidth}%
\caption{Calculated nonadiabatic coupled phonon-plasmon dispersion of
the $\Lambda_{1}$ modes of model OP for the optimally doped state of
LaCuO along the $\Lambda \sim (0, 0, 1)$ direction. The coupling modes
at $\Gamma$ (A$^{\text{LO}}_{2u}$ (ferro, na)) and at $Z$ (O$_z^Z$) are
shown as black dots ($\bullet$). The open dot ($\circ$) at $\Gamma$
indicates A$^{\text{TO}}_{2u}$ (ferro, ad). \\$-\!\!-$: $\Lambda_{1}$
modes; $-\,-$: free plasmon branch; $\cdots$: borderline
for damping; $-\cdot -$: steep $\Lambda_1$ branch in the adiabatic approximation.}\label{fig11}%
\end{minipage}%
\end{figure}
%%%%%%%%%%%%%%%%%%%%%%%%%%%%%%%%%%%%%%%%%%%%%%%%%%%%%%%%%%%%%%%%%%%%%%%%%%%%%%%%%%%%%%%%%
%%%%%%%%%%%%%%%%%%%%%%%%%%%%%%%%%%%%%%%%%%%%%%%%%%%%%%%%%%%%%%%%%%%%%%%%%%%%%%%%%%%%%%%%%

In Fig. \ref{fig11} we show the calculated nonadiabatic phonon-plasmon
dispersion of the $\Lambda_{1}$ modes along the $\Lambda \sim (0, 0,
1)$ direction. The full dots at $\Gamma$ and $Z$ represent the coupled
phonon-like (lower dot) and plasmon-like A$_{2u}$ FM (upper dot) and
the phonon-like (lower dot) and plasmon-like (upper dot) O$_z^Z$ mode,
respectively. The calculated longitudinal plasmon-like A$_{2u}$ FM at
18.41 THz is not far away from the experimental result for the
longitudinal A$_{2u}$ FM of LaCuO in the insulating state obtained with
INS at about 17.5 THz. Of course the large splitting between this mode
and the corresponding TO mode (open dot) accounts for the dominance of
the FM in the infrared response also in the insulating state
\cite{Ref14,Ref43}. It should be remarked that in order to obtain the
calculated dispersion curves the effect of the DF's is important. The
latter reduce the frequency of the plasmon as compared with the
calculation which takes only CF's into account. This effect is
exemplified in Table \ref{tab3} for the free-plasmon and the
plasmon-like modes at $\Gamma$ and $Z$.

\begin{table}
\begin{tabular}{c|ccccc}
    & $\Gamma$ & $Z$  & \hphantom{00000} & $\Gamma$ & $Z$    \\\hline
 CF & 14.09    & 9.83 & \hphantom{00000} & 20.45    & 16.99 \\
CFD & 12.78    & 9.51 & \hphantom{00000} & 18.41    & 13.86
\end{tabular}
\caption{Frequency in THz of the free $c$-axis plasmon at the $\Gamma$
and $Z$-point of the Brillouine zone (leftmost columns) as calculated
with model OP for the optimally doped state. The notation CF means that
only charge fluctuations are taken into account; CFD represents a
calculation including additionally anisotropic dipole fluctuations. The
rightmost columns gives the corresponding results for the plasmon-like
modes. Compare with Fig. \ref{fig11}.}\label{tab3}
\end{table}

From our preceding discussion of the coupled mode scenario along the
$\Lambda'$ and $\Lambda''$ direction it can be concluded that the
branch with the phonon-like FM at $\Gamma$ and the phonon-like
$O^{Z'}_{z}$ mode at $Z'$ converges to the steep $\Lambda_{1}$ branch
of Fig. \ref{fig11} in the $\Lambda$ direction when approaching the
adiabatic limit. Thereby the LO-TO splitting at $\Gamma$ is closed from
below and the phonon-like O$_z^Z$ mode reaches its adiabatic value. The
plasmon-like branch connecting the corresponding plasmon-like modes at
$\Gamma$ and $Z'$ rapidly leaves the phonon-spectrum as can be seen in
Fig. \ref{fig10}.

A comment is appropriate regarding the relation between electronic
correlations in the CuO plane beyond LDA and the possible size of the
nonadiabatic region which points to an interplay of nonadiabatic
effects around the $c$-axis due to the long-ranged Coulomb interaction
and strong short-ranged Coulomb repulsion responsible for correlation.
The latter leads in tendency to a reduction of the bandwidth and the
Fermi velocity of the quasi-particle (QP) excitations in the ($k_{x}$,
$k_{y}$)-plane as compared with DFT-LDA-like bandstructures
\cite{Ref57}. Such a reduction should increase the size of the
nonadiabatic region, because the electron dynamics in the CuO plane is
slowed down and the free-plasmon frequency is accordingly decreased. In
Ref. \onlinecite{Ref58} we have simulated a narrowing of the electronic
bandstructure within the 11BM by reducing empirically certain tight
binding parameters and as a result a slight increase of the
nonadiabatic region around the $c$-axis has been found. Thus, it can be
expected that characteristic properties of the QP in the CuO plane
related to strong correlation effects take influence on the dynamical
charge response perpendicular to the CuO layer nearby the $c$-axis.

Of course the position of the free-plasmon along the $c$-axis directly
depends on the strength of the interlayer coupling which may differ in
other cuprates from the situation found for LaCuO in the M31BM. As
already mentioned, we have studied in the past parametrically the
effect of a varying anisotropy in Refs. \onlinecite{Ref11,Ref12}. For a
very weak interlayer coupling the resulting nonadiabatic dispersion in
the metallic state along the $\Lambda$ direction virtually cannot be
distinguished from the dispersion of the (adiabatic) insulator and the
optical phonon modes remain completely unscreened.

As far as the pairing channel via coupled phonon-plasmon modes is
concerned it is interesting to remark that the total effective
interaction between the electrons via phonon-plasmon exchange is {\em
attractive} in the frequency range between the transverse A$_{2u}$ FM
and the longitudinal plasmon-like A$_{2u}$ FM at the $\Gamma$ point.
Thus, the very large splitting of this mode, see Fig. \ref{fig10} and
Fig. \ref{fig11}, is favourable for pairing and the role played in this
game by the incomplete dynamical screening of the (bare) polar Coulomb
interaction in the nonadiabatic region becomes evident. Note in this
context that in recent work \cite{Ref06} it is found that the
phonon-plasmon mechanism contributes constructively and significantly
to the superconductivity in the cuprates and that the long-range polar
electron-phonon interaction is important for pairing
\cite{Ref52,Ref53}.

%%%%%%%%%%%%%%%%%%%%%%%%%%%%%%%%%%%%%%%%%%%%%%%%%%%%%%%%%%%%%%%%%%%%%%%%%%%%%%%%%%%%%%%%%
%%%%%%%%%%%%%%%%%%%%%%%%%%%%%%%%%%%%%%%%%%%%%%%%%%%%%%%%%%%%%%%%%%%%%%%%%%%%%%%%%%%%%%%%%
\begin{figure}
\includegraphics[height=\linewidth,angle=90]{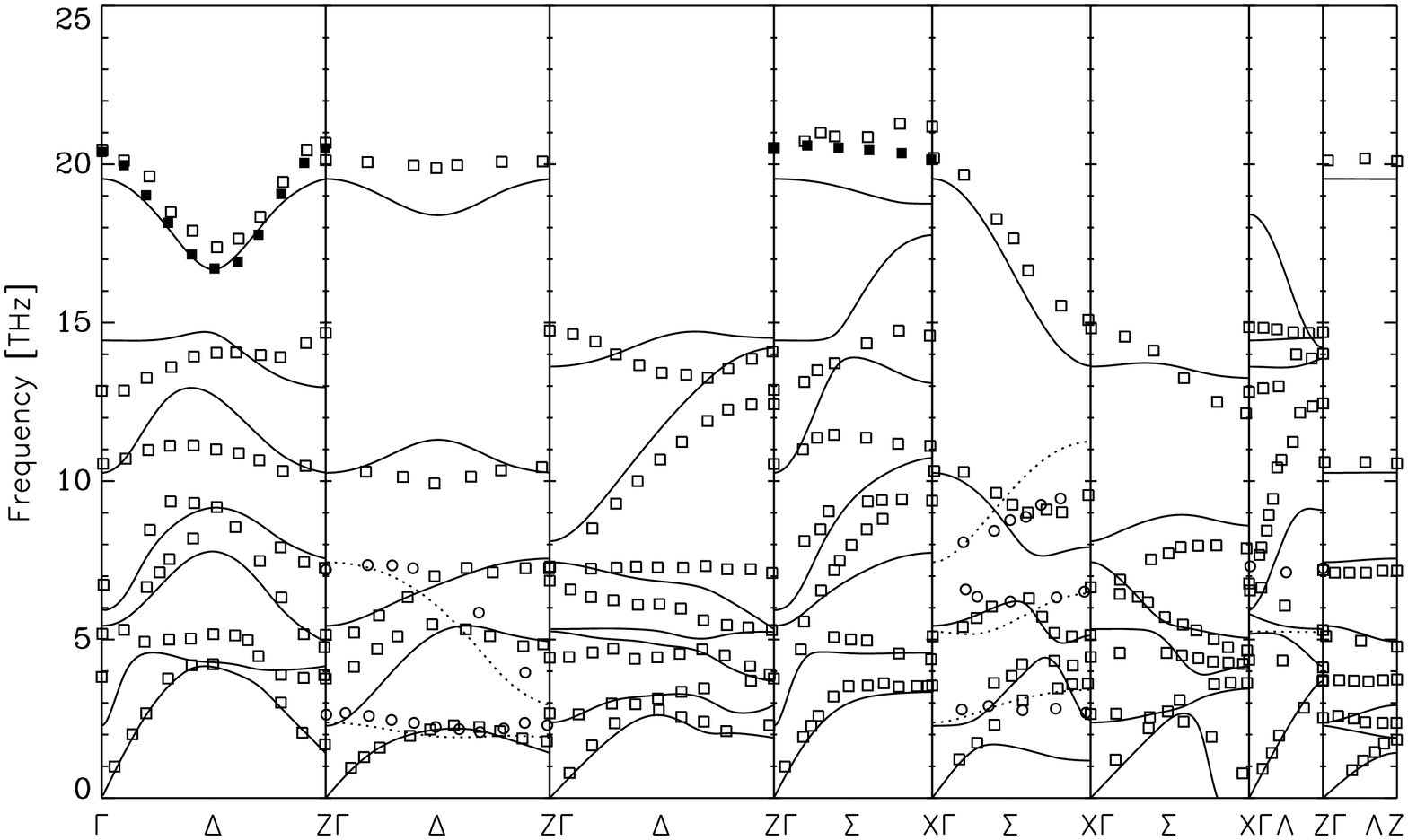}
\caption{Calculated nonadiabatic phonon dispersion of optimally doped
LaCuO in the main symmetry directions $\Delta \sim (1,0,0)$, $\Sigma
\sim (1, 1, 0)$ and $\Lambda \sim (0, 0, 1)$ within the M31BM (model
OP) for the proper polarization part. The various symbols representing
the experimental results from Refs. \onlinecite{Ref31,Ref32} indicate
different irreducible representation (ID's). Open symbols are for
La$_{1.9}$ Sr$_{0.1}$ CuO$_{4}$ and solid symbols for optimally doped
LaCuO. The arrangement of the panels from left to right according to
the different ID's is as follows: $|\Delta_{1}|\Delta_{2} (...,
\circ),$ $\Delta_{4} (-\!\!-, \Box)|\Delta_{3}|\Sigma_{1}|\Sigma_{2}
(\cdot\!\cdot\!\cdot, \circ)$, $\Sigma_{4} (-\!\!-, \Box)|$
$\Sigma_{3}|\Lambda_{1}(-\!\!-, \Box)$, $\Lambda_{2}
(\cdot~\cdot~\cdot~\cdot, \circ) |\Lambda_{3}|$.}\label{fig12}
\end{figure}
%%%%%%%%%%%%%%%%%%%%%%%%%%%%%%%%%%%%%%%%%%%%%%%%%%%%%%%%%%%%%%%%%%%%%%%%%%%%%%%%%%%%%%%%%
%%%%%%%%%%%%%%%%%%%%%%%%%%%%%%%%%%%%%%%%%%%%%%%%%%%%%%%%%%%%%%%%%%%%%%%%%%%%%%%%%%%%%%%%%

\begin{figure}
\includegraphics[height=\linewidth,angle=90]{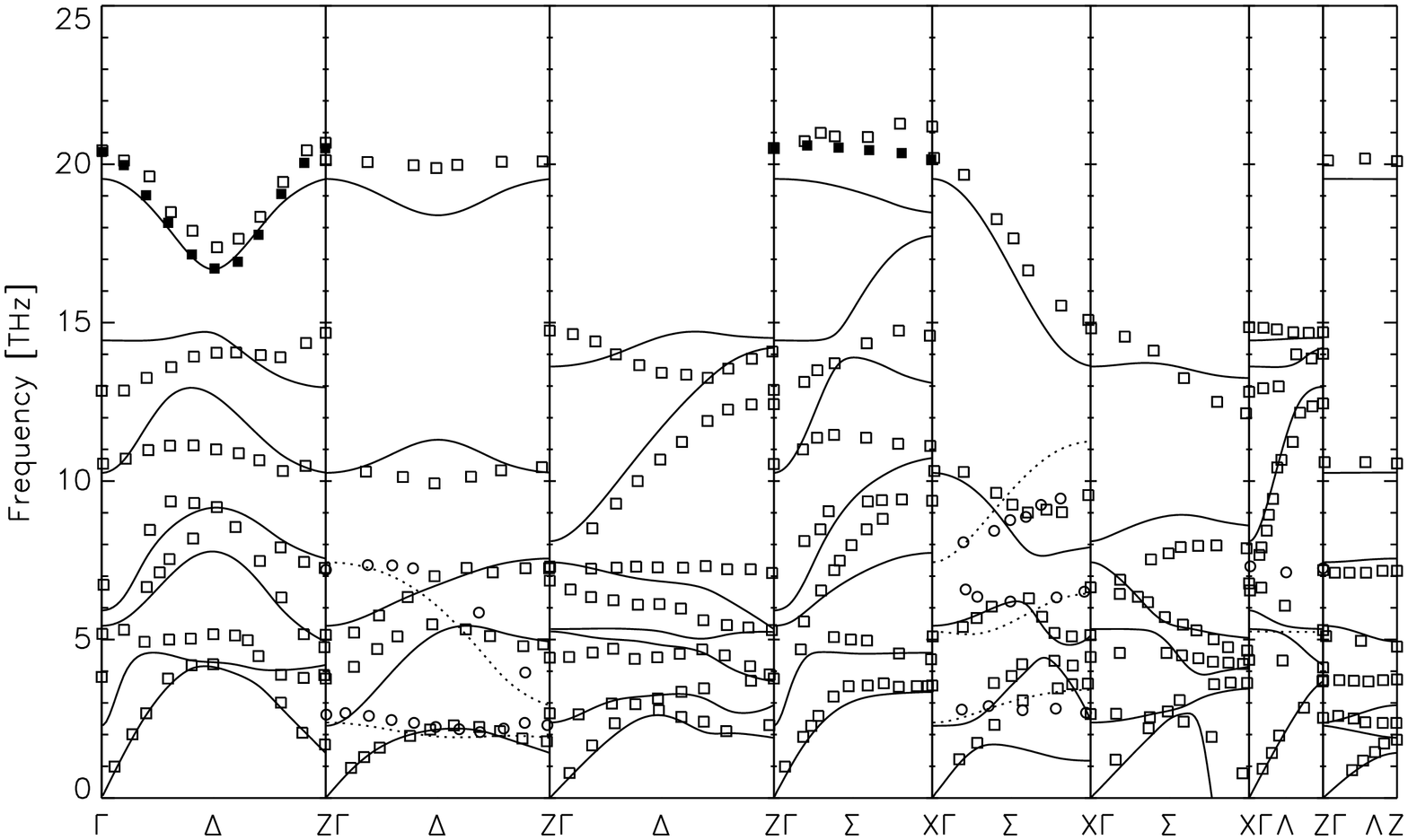}
\caption{Same as in Fig. \ref{fig12}, but calculated in the adiabatic
approximation.}\label{fig13}
\end{figure}

In Figs. \ref{fig12}, \ref{fig13} we compare our calculations of the
phonon dispersion of LaCuO along the main symmetry directions in the BZ
within the M31BM as obtained for the nonadiabatic and adiabatic case of
the proper polarization part, respectively. In Fig. \ref{fig12} the
small nonadiabatic region outside the $c$-axis where the adiabatic
approximation grossly fails has been excluded because it cannot be
resolved in this figure. Leaving aside the large differences of the
calculated dispersion for the $\Lambda_{1}$ modes as already discussed
virtually no changes can be observed when passing from the nonadiabatic
treatment to the adiabatic calculation. In particular this hold true
for the modes propagating in the CuO plane. The reason is the fast
electron dynamics in the CuO plane as compared to the dynamics
perpendicular to the plane. From all the modes propagating in the CuO
layer studied, the $O^{X}_{B}$ breathing mode, see Fig. \ref{fig06},
shows the largest deviation, namely a stiffening of 0.284 THz (9.47
cm$^{-1}$) for optimally doped LaCuO in the nonadiabatic treatment. We
attribute this increase of the frequency to a {\em dynamical} reduced
nesting effect with wavevectors around $X = \frac{\pi}{a} (1, 1, 0)$.
Such an interpretation is consistent with the calculated FS in Fig.
\ref{fig04}(a) and the maximum found at the ${X}$-point for the reduced
electronic polarizability function $\Pi_{0} (\vc{q})$ from Eq.
\eqref{Eq26}. The change of the FS topology upon doping from hole-like
to electron-like and the related change of the nesting structures in
Fig. \ref{fig04} leads in the calculations to a reduction of the
nonadiabatic stiffening of $O^{X}_{B}$ (model OD1: 0.193 THz (6.44
cm$^{-1}$), model OD2: 0.059 THz (1.97 cm$^{-1}$)).

%%%%%%%%%%%%%%%%%%%%%%%%%%%%%%%%%%%%%%%%%%%%%%%%%%%%%%%%%%%%%%%%%%%%%%%%%%%%%%%%%%%%%%%%%
%%%%%%%%%%%%%%%%%%%%%%%%%%%%%%%%%%%%%%%%%%%%%%%%%%%%%%%%%%%%%%%%%%%%%%%%%%%%%%%%%%%%%%%%%
%%%%%%%%%%%%%%%%%%%%%%%%%%%%%%%%%%%%%%%%%%%%%%%%%%%%%%%%%%%%%%%%%%%%%%%%%%%%%%%%%%%%%%%%%
%%%%%%%%%%%%%%%%%%%%%%%%%%%%%%%%%%%%%%%%%%%%%%%%%%%%%%%%%%%%%%%%%%%%%%%%%%%%%%%%%%%%%%%%%

A possible effect of the nonadiabatic charge response on the Raman
modes in LaCuO could be quite interesting because very recently giant
nonadiabatic effects in layered metals such as graphite intercalated
compounds, which are three-dimensional metals with a considerable
anisotropy along one direction, have been predicted in Ref.
\onlinecite{Ref59}. Contrarily, in our calculations for LaCuO
practically no nonadiabatic  effects have been found for the Raman
modes.

Finally, a comment on the calculated unstable mode in Figs.
\ref{fig12}, \ref{fig13} at the $X$-point should be made. This mode is
the so called tilt mode. Freezing of this distortion points correctly
to the experimentally observed structural phase transition from the
high-temperature tetragonal (HTT) to the low-temperature orthohombic
(LTO) structure. In earlier calculations for the LTO structure we have
found within the ab initio RIM two stable tilt modes with different
frequencies in reasonable agreement with the experiment \cite{Ref12}.
From these calculations it can be concluded that the structural phase
transition is essentially driven by the strong long-ranged Coulomb
interactions between the ions in the material which again emphasizes
the important role of ionic binding for the cuprates.

\subsection{Phonon-plasmon induced charge response and selfconsistent changes of the potential}
In previous work \cite{Ref12,Ref60,Ref61} we have made explicit the
charge response by calculating in adiabatic approximation the
phonon-induced charge density redistribution for the presumably generic
phonon anomalies $\Delta_{1}/2$ (half-breathing mode) and $O^{X}_{B}$
(planar-breathing mode), essentially polarized in the CuO plane and
also for the O$_z^Z$ mode \cite{Ref28} polarized perpendicular to the
plane, see Fig. \ref{fig06} for the displacement patterns of these
oxygen-bond-stretching modes (OBSM). We have shown in Sec. B that
$\Delta_{1}/2$ and $O^{X}_{B}$ can be treated in the adiabatic limit,
but this is of course not true for O$_z^Z$ as a specific important mode
from the nonadiabatic region of charge response.

As suggested in Ref. \onlinecite{Ref62} specific modes (in particular
$\Delta_{1}/2$) with a strong electron-phonon coupling can generate an
\textit{overscreening} of the intersite Coulomb interaction due to a
phonon-induced charge transfer between Cu and O$_{xy}$. This strong
coupling has been proposed in \cite{Ref62} to form a basis for the
phonon mechanism of high-temperature superconductivity.

The anomalous softening of the OBSM leading as in case of
$\Delta_{1}/2$ to a decreasing dispersion when starting at the $\Gamma$
point, see Figs. \ref{fig12},\ref{fig13}, is a result of
\textit{overscreening} at shorter wave lengths of the changes of the
Coulomb potential generated by the motion of the ions. This has been
shown by our calculations to be due to nonlocally excited ionic CF's
localized at the Cu and O$_{xy}$ sites generating a dynamic charge
ordering in form of localized stripes of alternating sign in the CuO
layer. Such induced charge patterns also have been calculated quite
recently for $n$-doped NdCuO \cite{Ref29}. The importance for the
phonon anomalies $\Delta_{1}/2$ and $O^{X}_{B}$ of the repulsive
on-site Coulomb interaction $U_{d}$ as a relevant physical parameter of
the cuprates deciding about the correlations in these materials has
been predicted already in Ref. \onlinecite{Ref39} and investigated in
more detail in Refs. \onlinecite{Ref28,Ref60}. Briefly, a large $U_{d}$
suppresses double occupancy of the Cu3d orbital and thus Cu3d CF's
which are crucial for the softening of $O^{X}_{B}$ and $\Delta_{1}/2$.

For the $\Delta_{1}/2$ mode the charge stripes point along the CuO
bonds and for $O^{X}_{B}$ along the diagonals of the CuO plane.
Moreover, in Ref. \onlinecite{Ref12} we have pointed out by means of
the OBSM, how antiferromagnetic correlations may favour the CF's in
these modes and vice versa that nonlocal EPI is expected to generate
antiferromagnetic spin-fluctuations (SF's) via the phonon-induced CF's
between the Cu ions. This interplay between charge- and spin-degrees of
freedom should die out in the overdoped state because of the vanishing
of the antiferromagnetic correlations.

%%%%%%%%%%%%%%%%%%%%%%%%%%%%%%%%%%%%%%%%%%%%%%%%%%%%%%%%%%%%%%%%%%%%%%%%%%%%%%%%%%%%%%%%%
%%%%%%%%%%%%%%%%%%%%%%%%%%%%%%%%%%%%%%%%%%%%%%%%%%%%%%%%%%%%%%%%%%%%%%%%%%%%%%%%%%%%%%%%%

\begin{figure*}%
\begin{minipage}{0.5\linewidth}%
  \centering%
  \subfigure[]{\includegraphics[height=7cm, width=7cm]{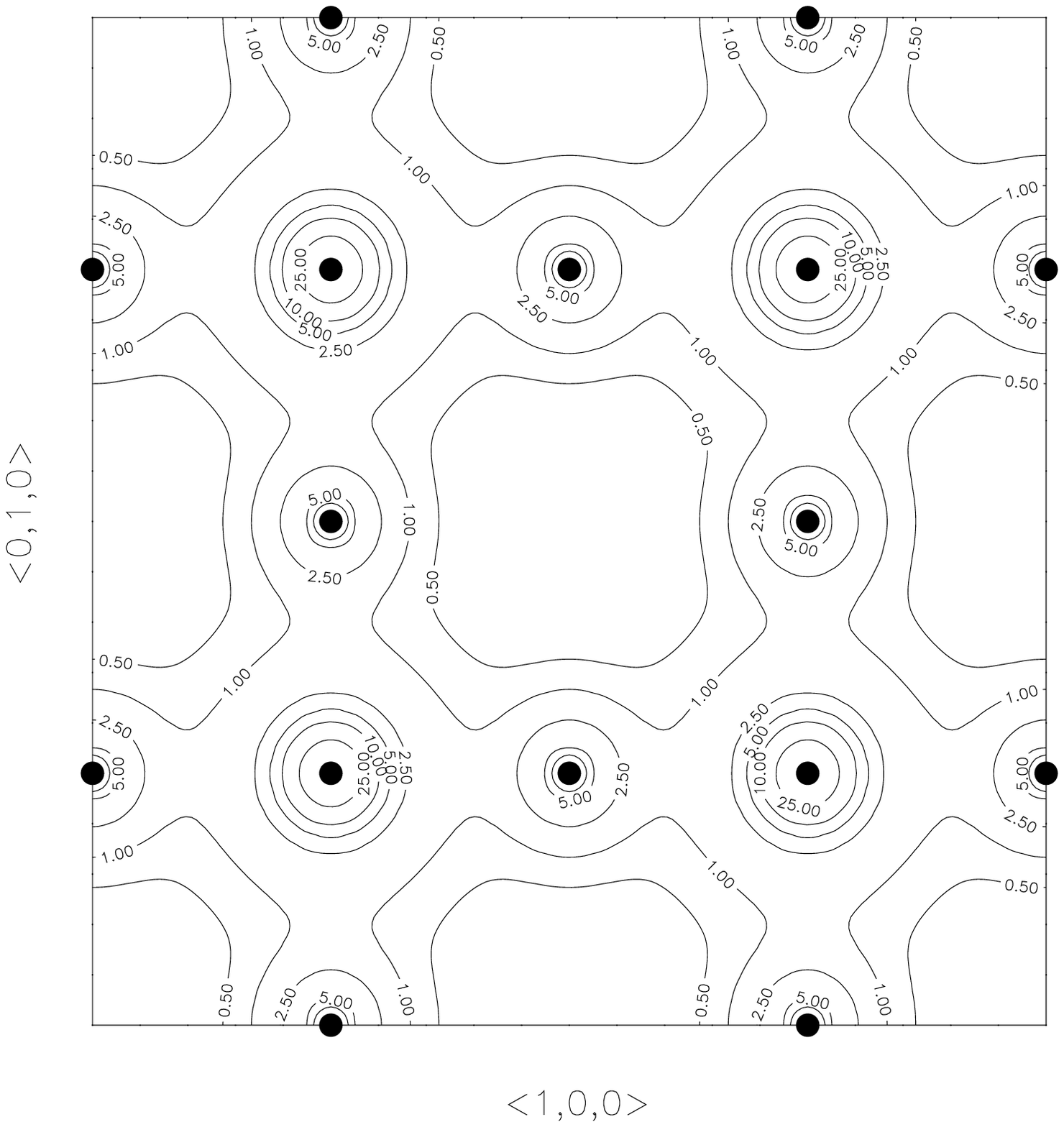}}%
\end{minipage}%
\begin{minipage}{0.5\linewidth}%
  \centering%
  \subfigure[]{\includegraphics[height=7cm, width=7cm]{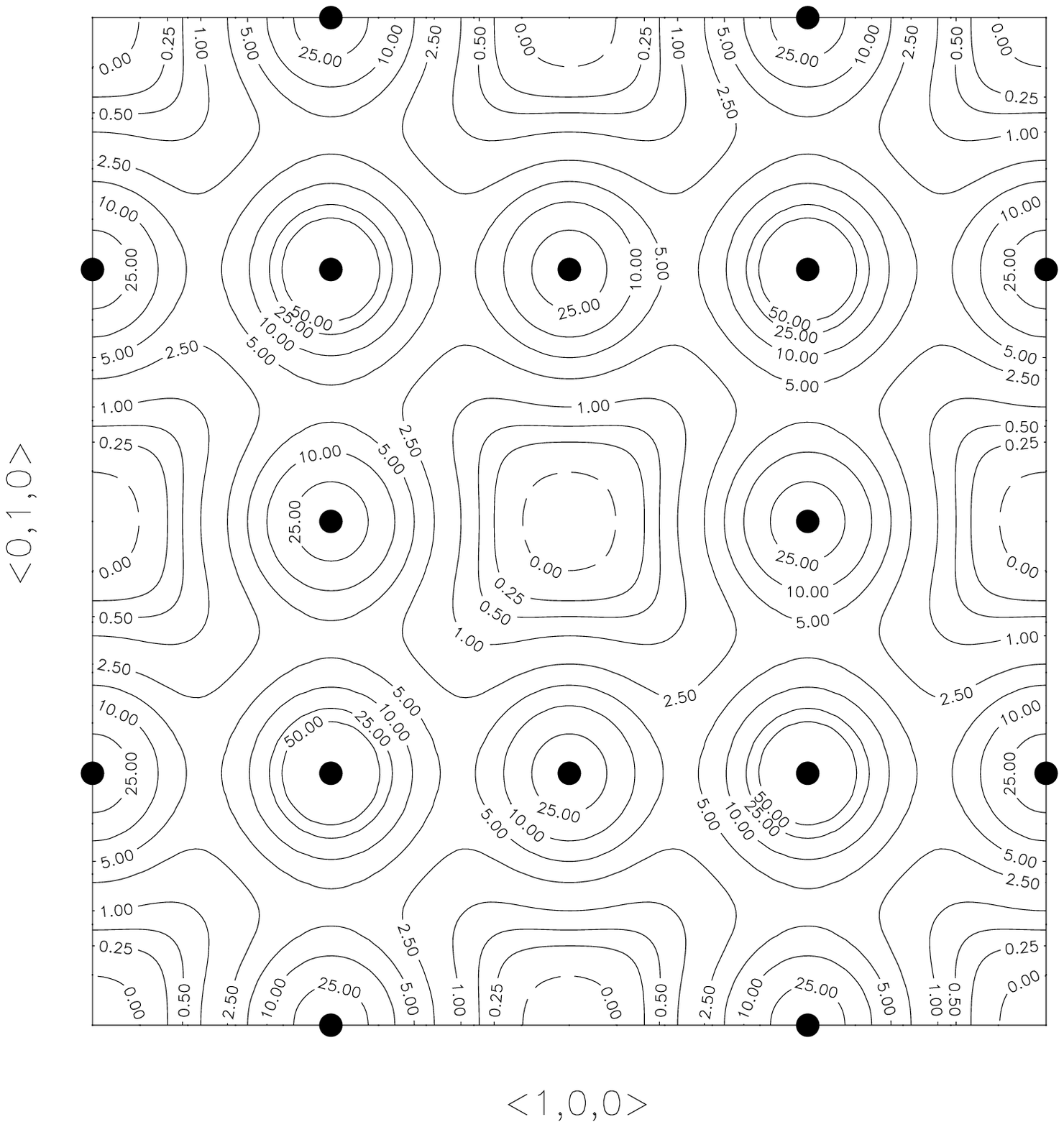}}%
\end{minipage}%

\begin{minipage}{0.5\linewidth}%
  \centering%
  \subfigure[]{\includegraphics[height=7cm, width=7cm]{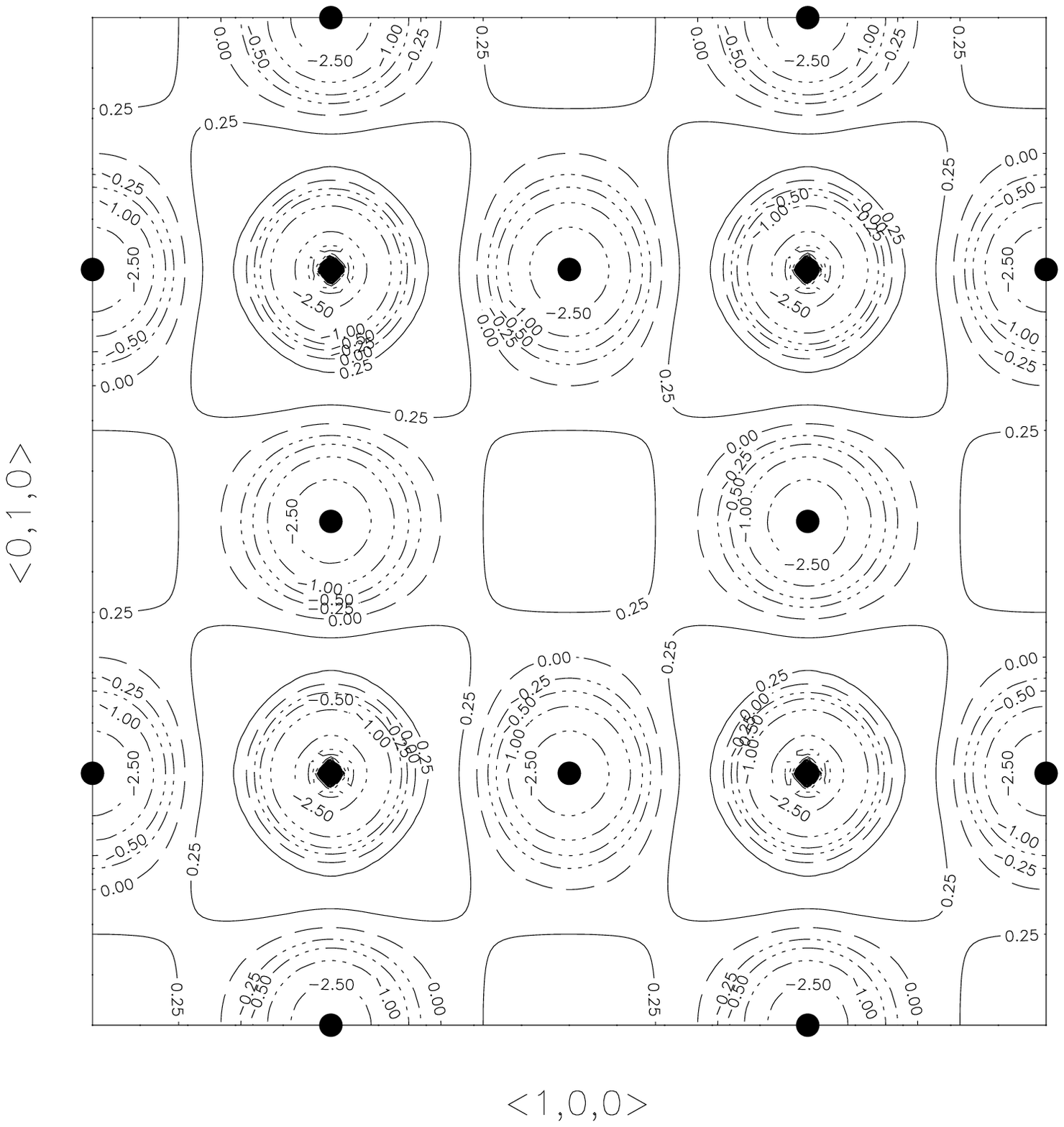}}%
\end{minipage}%
%\hspace{0.1\linewidth}%
\begin{minipage}{0.5\linewidth}%
  \caption{Contour plot in the CuO plane of the nonlocal part of the
displacement-induced charge density redistribution for the O$_z^Z$ mode
in LaCuO calculated according to Eq. \eqref{Eq30} in adiabatic
approximation for the M31BM (model OP) for the proper polarization part
(a). (b) same as in (a) but for the phonon-like O$_z^Z$ mode. (c) same
as (a) but for the plasmon-like O$_z^Z$ mode. The units are in
$10^{-4}$ $e^{2}/a^{3}_{B}$. The phase of O$_z^Z$ is as shown in Fig.
\ref{fig06}. Full lines ($-\!\!-$) mean that electrons are accumulated
in the corresponding region of space and the broken-dot lines ($-\cdots
-$) indicate regions where the electrons are
pushed away.}\label{fig14}%
\end{minipage}%
\end{figure*}%

%%%%%%%%%%%%%%%%%%%%%%%%%%%%%%%%%%%%%%%%%%%%%%%%%%%%%%%%%%%%%%%%%%%%%%%%%%%%%%%%%%%%%%%%%
%%%%%%%%%%%%%%%%%%%%%%%%%%%%%%%%%%%%%%%%%%%%%%%%%%%%%%%%%%%%%%%%%%%%%%%%%%%%%%%%%%%%%%%%%

The induced charge redistribution for O$_z^Z$ calculated in the
adiabatic limit has been addressed shortly in Sec. B and will be
compared explicitly in Fig. \ref{fig14} with the corresponding results
as obtained in the nonadiabatic approach. So far we can say that the
phonon anomalies $\Delta_{1}/2$ and $O^{X}_{B}$ reflect the importance
of the short-ranged part of the Coulomb interaction ($U_d$) while
O$_z^Z$ is specifically governed by the long-ranged part.

Before proceeding with the investigation of the charge response in the
optimally to overdoped metallic state of LaCuO we will include for
completeness a short summary of our approach to phonon dynamics and
charge response in the insulating  and underdoped state of the
cuprates. A detailed discussion can be found in Refs.
\onlinecite{Ref12,Ref28,Ref29,Ref60,Ref63}. In this work also the
strong \textit{doping dependence} of the OBSM phonon anomalies has been
studied.

In order to describe these states of the cuprates we have proposed a
microscopic modeling which takes into account the enhanced correlation
of the Cu3d orbitals in these ''phases'' via a
compressibilty-incompressibility transition for the Cu3d state in terms
of rigorous sum rules for the proper polarization part
$\Pi_{\kappa\kappa'} (\vc{q})$ in the longwavelength-limit. An
interesting aspect of the partial incompressibility of the Cu3d states
is that in case pairing correlations (pre-pairing) would exist above
$T_{C}$ the reduced local particle number fluctuations related to the
orbital incompressibility of Cu3d due to the large on-site repulsion
would correlate with enhanced phase fluctuations according to the
particle number phase uncertainity relation and consequently a
collective onset of phase coherence is suppressed. This in turn would
stabilize a pre-paired state above $T_{C}$.

In Refs. \onlinecite{Ref29,Ref63} also the relation of the
superconducting gap and the pseudogap as well as the possibility of
Fermi surface transformations observed in the normal state of the
cuprates is discussed within the model. Moreover, the modeling has been
extended to $n$-doped cuprates and the \textit{electron-hole asymmetry}
is simulated via an orbital selective incompressibility-compressibility
transition of the O2p orbital in contrast to the $p$-doped case.

We now return to the investigation of the phonon-induced nonadiabatic
charge response in optimally doped LaCuO represented by the model OP.
The nonlocal, nonrigid part of the charge response related to the
nonlocal EPI effects as excited by a phonon with wavevector $\vc{q}$
and polarization $\sigma$ is given by
\begin{equation} \label{Eq30}
\delta \rho_{n} (\vc{r}, \vc{q}\sigma) = \sum\limits_{\vc{a}, \kappa}
\delta \zeta^{\vc{a}}_{\kappa} (\vc{q},\sigma) \rho_{\kappa} (\vc{r} -
\vc{R}_{\kappa}^{\vc{a}}) .
\end{equation}
The CF's $\delta \zeta^{\vc{a}}_{\kappa}$ are obtained from Eq.
\eqref{Eq19} and the form-factors $\rho_{\kappa}$ from Eq. \eqref{Eq2}.

In the contour plots of Fig. \ref{fig14} the displacement induced
rearrangement of the charge density according to Eq. \eqref{Eq30} is
shown for the O$_z^Z$ mode. The calculation has been performed  for the
model OP. Fig. \ref{fig14}(a) displays the result for the adiabatic
approximation and Fig. \ref{fig14}(b) and Fig. \ref{fig14}(c) pictures
the nonadiabatic result for the phonon-like and plasmon-like mode
O$_z^Z$ mode, respectively. The phase of the vibration of the ions is
as in Fig 6, i.e. the apex oxygens move away from the CuO layer. The
full lines ($\delta \rho_{n}
> 0)$ indicate that electrons are accumulated in the associated region
of space. The broken lines $(\delta \rho_{n} < 0)$ mean that electrons
are depleted in this region.

%%%%%%%%%%%%%%%%%%%%%%%%%%%%%%%%%%%%%%%%%%%%%%%%%%%%%%%%%%%%%%%%%%%%%%%%%%%%%%%%%%%%%%%%%
%%%%%%%%%%%%%%%%%%%%%%%%%%%%%%%%%%%%%%%%%%%%%%%%%%%%%%%%%%%%%%%%%%%%%%%%%%%%%%%%%%%%%%%%%
\begin{figure}
\includegraphics[width=\linewidth]{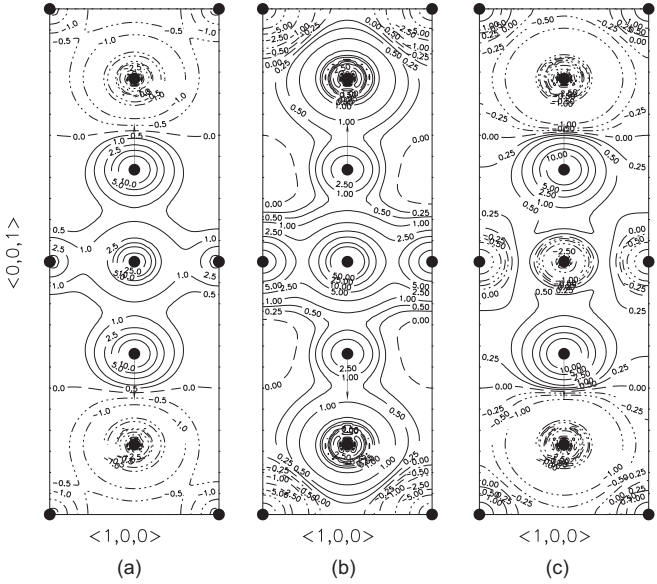}
\caption{Contour plot perpendicular to the CuO plane of the nonlocal
part of the displacement-induced charge density redistribution for the
O$_z^Z$ mode in LaCuO calculated from Eq. \eqref{Eq30} in adiabatic
approximation for the M31BM (model OP) (a). (b) same as (a) but for the
phonon-like O$_z^Z$ mode. (c) same as (a) but for the plasmon-like
O$_z^Z$ mode. The units and the meaning of the line types as in Fig.
\ref{fig14}. The arrows denote the displacements of the apex oxygen
ions.}\label{fig15}
\end{figure}
%%%%%%%%%%%%%%%%%%%%%%%%%%%%%%%%%%%%%%%%%%%%%%%%%%%%%%%%%%%%%%%%%%%%%%%%%%%%%%%%%%%%%%%%%
%%%%%%%%%%%%%%%%%%%%%%%%%%%%%%%%%%%%%%%%%%%%%%%%%%%%%%%%%%%%%%%%%%%%%%%%%%%%%%%%%%%%%%%%%

As already mentioned in Sec. B and displayed explicitly in Fig. 14(a)
in the adiabatic approximation O$_z^Z$ (12.96 THz) generates CF's at
the Cu and O$_{xy}$ ions of the same sign in the whole CuO layer. In
the adjacent layers the induced CF's are the same in magnitude but of
opposite sign, compare with Fig. \ref{fig15}(a). Thus, O$_z^Z$
activates an interlayer charge transfer which is instantaneous in
adiabatic approximation. Such a nonlocally excited screening process is
very effective for screening the changes of the Coulomb interaction
generated by the displacement of the apex oxygen ions and explains
accordingly the large softening of O$_z^Z$ during the insulator-metal
transition.

Fig. \ref{fig14}(b) records our calculated results for the
nonadiabatically induced CF's of the phonon-like O$_z^Z$ mode in the
CuO plane  at 9.09 THz. In the nonadiabatic regime the electrons cannot
follow instantaneously the movement of the ions. The now dynamically
excited CF's  at Cu and O$_{xy}$ have the same sign in the CuO layer.
The sign is as in the adiabatic case because the frequency of the free
plasmon at $Z$ is (slightly) higher (9.5 THz, see Table \ref{tab3})
than the phonon-like O$_z^Z$ mode. The magnitude of the CF's is
considerably larger in the nonadiabatic case, see also Table
\ref{tab4}.

The interlayer CT is now of dynamic nature because of the dynamical
screening of the Coulomb interaction and consequently a coupled
collective phonon-like excitation propagating along the $c$-axis is
created leading to the \textit{correlated} charge rearrangement between
the layers due to the \textit{coherent} dynamics as shown in the
snapshot of Fig. \ref{fig15}(b). Comparing with the adiabatic limit
besides the strongly enhanced charge response, which explains the lower
frequency, the sign of the CF's at the La ion has changed as compared
with the adiabatic calculation. This correlates well with the change of
sign of the displacement amplitude for La, compare with Table
\ref{tab2}. On the other hand, the sign of the CF's at the La for the
plasmon-like mode is the same as in the adiabatic case in agreement
with the same sign of the displacement of La in both calculations.

The coupled plasmon-like O$_z^Z$ mode at 13.86 THz has a higher
frequency than the free plasmon and apparently the charge response runs
out of phase, see Fig. \ref{fig14}(c). This results in an
\textit{antiscreening} consistent with the higher mode frequency. The
sign of the CF's at Cu and O$_{xy}$ is not only opposite to that in the
adiabatic and phonon-like case but also smaller in magnitude (Table
\ref{tab4}). Fig. \ref{fig14}(c) together with Fig. \ref{fig15}(c) give
a snapshot of the collective plasmon-like interplane charge excitations
being out of phase with the lattice and also strongly reduced in
magnitude as compared with the phonon-like excitation displayed in Fig.
\ref{fig14}(b) and Fig. \ref{fig15}(b), respectively.

Concerning $c$-axis transport in a highly anisotropic metal it is
interesting to note that in a simplified model treatment the
possibility has been pointed out that a strong coupling between
electrons and a bosonic mode propagating and polarized in the
$c$-direction, like O$_z^Z$, can lead to a metallic to nonmetallic
crossover in the $c$-axis resistivity as the temperature is increased
with no corresponding feature in the $xy$ plane properties
\cite{Ref64}.

%%%%%%%%%%%%%%%%%%%%%%%%%%%%%%%%%%%%%%%%%%%%%%%%%%%%%%%%%%%%%%%%%%%%%%%%%%%%%%%%%%%%%%%%%
%%%%%%%%%%%%%%%%%%%%%%%%%%%%%%%%%%%%%%%%%%%%%%%%%%%%%%%%%%%%%%%%%%%%%%%%%%%%%%%%%%%%%%%%%
\begin{figure}
\includegraphics[width=\linewidth]{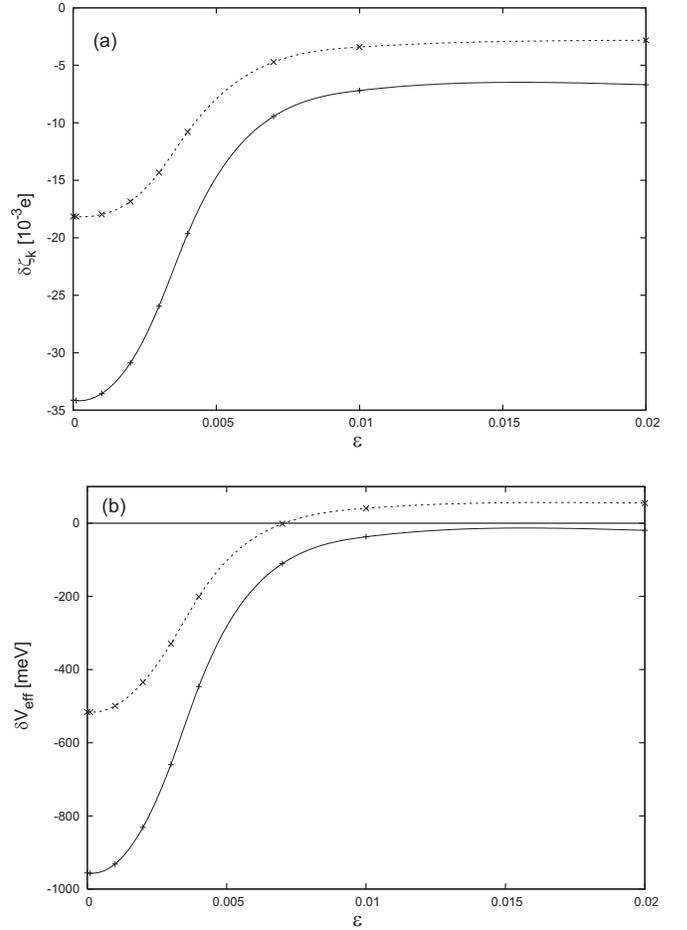}
\caption{Calculated displacement-induced charge-fluc\-tua\-tions
$\delta \zeta_{\kappa} (\vc{q}, \sigma, \omega)$ (Eq. \eqref{Eq19}) (a)
and corresponding orbital averaged changes of the crystal potential
$\delta V_{\kappa} (\vc{q}, \sigma, \omega)$ (Eq. \eqref{Eq22}) (b) for
the phonon-like O$_z^{Z'}$ mode along the $\Delta'$ direction
$\left(\varepsilon \frac{2\pi}{a}, 0 , \frac{2\pi}{c}\right)$,
$\varepsilon \in (0.00, 0.02)$. Cu3d ($-\!\!\!-$) and $O_{xy}$2p
($\cdots$) orbital degrees of freedom are considered. The calculations
have been performed for model OP.}\label{fig16}
\end{figure}
%%%%%%%%%%%%%%%%%%%%%%%%%%%%%%%%%%%%%%%%%%%%%%%%%%%%%%%%%%%%%%%%%%%%%%%%%%%%%%%%%%%%%%%%%
%%%%%%%%%%%%%%%%%%%%%%%%%%%%%%%%%%%%%%%%%%%%%%%%%%%%%%%%%%%%%%%%%%%%%%%%%%%%%%%%%%%%%%%%%

In order to get an impression for the displacement induced changes of
the CF's $\delta \zeta_{\kappa}^{\vc{a}}(\vc{q}, \sigma)$ according to
Eq. \eqref{Eq19} and the corresponding induced orbital averaged changes
of the potential felt by the electrons $\delta V_{\kappa}^{\vc{a}}
(\vc{q}, \sigma)$ from Eq. \eqref{Eq22}, when passing from the region
with a nonadiabatic dynamic charge response to the region with
practically static adiabatic charge response, we display in Figs.
\ref{fig16}(a,b) our calculated results for the phonon-like
$O^{Z'}_{z}$ mode along the $\Delta'$ direction, compare with Fig.
\ref{fig07}(a). The calculations have been carried out for the
optimally doped case using model OP and for the most important Cu3d and
O$_{xy}$2p orbitals, respectively. Similar as for the phonon
frequencies in Fig. \ref{fig07}(a) $\delta \zeta_{\kappa}$ and $\delta
V_{\kappa}$ converge beyond $\varepsilon \approx 0.01$ to their values
in the adiabatic limit. Taking for example the calculated numbers at
$\varepsilon = 0.00$ and $\varepsilon = 0.02$, respectively, we
recognize that the nonlocally excited CF's and in particular the
nonlocally induced orbital averaged changes of the potential being a
measure of the long-ranged polar coupling of the electrons and phonons
are very strongly enhanced in the nonadiabatic region of phase space
around the $c$-axis.

Additionally, in Table \ref{tab4} we summarize and compare the
calculated orbital selective values of $\delta \zeta_{\kappa}$ and
$\delta V_{\kappa}$ for Cu3d and O$_{xy}$2p of the phonon anomalies
$\Delta_{1}/2$, $O^{X}_{B}$ and O$_z^Z$, respectively. We extract
already strong coupling of the adiabatic modes $\Delta_{1}/2$,
$O_{B}^{X}$ and of course the strong enhancement for the phonon-like
O$_z^Z$ mode. This again points to the fact that besides the
short-ranged part of the Coulomb interaction $U_{d}$ being important
for $\Delta_{1}/2$ and $O^{X}_{B}$, the strong long-ranged, nonlocal
polar Coulomb interaction must play an important role in the normal and
superconducting state of the doped cuprates.

\begin{table*}
\begin{tabular}{c|cc|cc|cc|cc|cc|cc|cc|}
&  \multicolumn{2}{|c|}{O$_z^Z$(ad)} &
\multicolumn{2}{|c|}{O$_z^Z$(na,ph)} &
\multicolumn{2}{|c|}{O$_z^Z$(na,pl)} &
\multicolumn{2}{|c|}{O$^X_\text{B}$(ad)} &
\multicolumn{2}{|c|}{$\Delta_1/2$(ad)} &
\multicolumn{2}{|c|}{O$^X_\text{B}$(na)} &
\multicolumn{2}{|c|}{$\Delta_1/2$(na)} \\\hline
 & Cu3d & O2p & Cu3d & O2p & Cu3d & O2p & Cu3d & O2p & Cu3d & O2p & Cu3d & O2p & Cu3d & O2p \\\hline
$\delta\zeta_\kappa$ & -6.59 & -2.63 & -34.12 & -18.13 & 1.32 & 2.13 &
-16.45 & 0 & -14.76 & -1.44 & -16.95 & 0 & -14.76 & -1.44\\
$\delta V_\kappa$ & -15.77 & 58.67 & -955.98  & -515.98 & 241.34 &
207.04 & -93.92 & 0 & -73.06 & 55.14 & -87.20 & 0 & -73.13 & 55.08
\end{tabular}
\caption{Magnitudes of the charge fluctuations $\delta\zeta_{\kappa}$
(Eq. \eqref{Eq19}) and orbital averaged changes of the selfconsistent
changes of the crystal potential $\delta V_{\kappa}$ (Eq. \eqref{Eq22})
of the OBSM as calculated with model OP for the Cu3d and the O2p
orbital in the CuO plane. $\delta\zeta_{\kappa}$ is given in units of
$10^{-3}$ particles and $\delta V_{\kappa}$ in units of meV. na:
nonadiabatic; ad: adiabatic; ph: phonon-like; pl: plasmon-like. The
phase of the displacements for the OBSM is as shown in Fig.
\ref{fig06}. $\delta \zeta_{\kappa} < 0$ means an accumulation of
electrons in the corresponding orbital. $\delta V_{\kappa} > 0$
indicates that the region around the ion is attractive for
electrons.}\label{tab4}
\end{table*}

In context with the very strong coupling of the {\em doping dependent}
phonon-like O$_z^Z$ mode which is at 9.09 THz (37.6 meV) in model OP
for the optimally doped state a special remark should be made. This
mode involves momentum transfer between antinodal regions of the FS
(similar as $O^{X}_{B}$ or the resonant magnetic mode), however, with a
drastically enhanced coupling strength as compared to $O_{B}^{X}$, see
Table \ref{tab4}. Consequently, this mode should contribute to the
electron self-energy via electron-phonon coupling in the antinodal
region and possibly also to the antinodal pseudogap in the normal
state. In the superconducting state even an increased contribution to
the self-energy via O$_z^Z$ can be expected because of the large
density of states enhancement in these regions due to the opening of a
gap with $d$-wave symmetry \cite{Ref65}. Moreover, below $T_{C}$ we
will have a suppression of the low scattering rate of the electrons due
to the opening of a gap and additionally below the finite energy of the
phonon-like mode.

This might be helpful to understand the dramatic change in lineshape of
the antinodal spectra seen in ARPES, where a sharp quasi-particle peak
develops at the lowest binding energies followed by a dip and a broader
hump. In the current literature these features are explained by a
finite energy collective bosonic mode which mostly is identified as the
magnetic resonance mode \cite{Ref66,Ref67}. On the other hand, the
enhanced coherence of the quasi-particles at lower temperature in the
superconducting state is simultaneously favourable for the existence of
a phonon-plasmon scenario with the plasmon-like mode as another
possibility for the collective bosonic mode.

From our calculations damping of the plasmon due to electron-hole decay
is less important because only intraband excitations contribute in case
of the simple BS near the Fermi level. So, the phonon-like mode should
be an important low-energy bosonic excitation mediating an attractive
interaction. Comparing with the isostructural low-temperature
superconductor SrRuO these relations can be expected to change
considerably because of the complex BS around the Fermi energy where
three FS sheets exist \cite{Ref68,Ref69,Ref70}. In such a situation a
strong $\vc{q}$-dependent  plasmon damping by low-lying interband
transitions can be expected in the collisionless regime and the
$c$-axis plasmon might cease to be a well defined excitation. Note,
that in simple metals is has been shown that interband transitions are
the dominant processes in the damping of plasmons \cite{Ref71}.
Additional damping due to interactions between electrons and with other
degrees of freedom enhances the damping still further and broadens the
energy levels. This leads to a narrowing of the interband gaps and to
the fact that the decay via interband transitions would even be
promoted. Altogether, it seems likely that in contrast to LaCuO in the
low-temperature superconductor SrRuO the $c$-axis phonon-plasmon modes
are not well defined. On the other hand, the existence of these modes
in the cuprates could point to their relevance for high-temperature
superconductivity.

In case of cuprate superconductors with more than one CuO layer a
qualitatively new possibility may arise as candidate for the finite
energy collective bosonic mode. In these compounds additionally to the
low-lying plasmon (intraband plasmon) of the single layer compounds
discussed so far which results from a weak but non vanishing $k_z$
dispersion of the QP's an interband plasmon with a low energy can
exist. The latter emerges at vanishing $k_z$ dispersion from a finite
but small intra-layer coupling, e.g. between the two CuO layers in a
double layer system. Such a possibility has been investigated
qualitatively for YBaCuO within a simplified model approach, see Refs.
\onlinecite{Ref12,RefHoffmann} analogously to the model for LaCuO in
Refs. \onlinecite{Ref11,Ref12}.

In case of YBaCuO a 22BM is used and then the degeneracy of the BS is
lifted by introducing parametrically an intra-bilayer coupling. The
latter is varied to simulate a corresponding bilayer splitting.
According to these calculations a sufficiently weak bilayer coupling
generates a nearly dispersionless low-lying phonon-like bond bending
branch polarized along the $c$-axis with a phonon-like
$A_g^Z(\text{O}_{23}\upuparrows)$ mode at the $Z$-point and a
longitudinal phonon-like FM ($B_{1u}^\Gamma(\text{O}_{23}\upuparrows)$)
at the $\Gamma$-point. The notation $(\text{O}_{23}\upuparrows)$ means
that the $\text{O}_{2}$ and $\text{O}_{3}$ oxygen ions in the CuO layer
are vibrating in phase. With increasing bilayer-coupling the frequency
of the phonon-like $A_g^Z(\text{O}_{23}\upuparrows)$ varies between
around 35 meV at weak coupling and 46 meV in the adiabatic limit.

The calculated phonon-induced charge rearrangement of
$A_g^Z(\text{O}_{23}\upuparrows)$ in YBaCuO is similar to that of
O$^Z_z$ in LaCuO as far as a double layer of YBaCuO is concerned. The
CF's have the same sign in the CuO plane but are of opposite sign in
the two CuO planes in the double layer. This is energetically
favourable because of the attractive Coulomb interaction at short
distance between the CF redistribution of the two layers. So, this CF
pattern also explains the strong renormalization of the frequency of
this mode as compared with a calculation where the CF's and the related
coupling within de bilayer is ignored (61 meV) and DF's only are
considered\cite{RefHoffmann}.

Concerning the infrared response the weak bilayer-coupling only leads
to a partial screening of the long-ranged Coulomb interaction and thus
to a corresponding fractional reduction of the LO-TO splittings, i.e.
the latter are still present consistent with the measured optical
$c$-axis spectra in YBaCuO, see e.g. Refs.
\onlinecite{RefSchuetzmann,RefHenn}. However, the splittings are
gradually closed from below by admitting additionally to the
intra-bilayer coupling an inter-bilayer coupling in the model leading
to a non vanishing $k_z$ dispersion and, finally, to the complete
metallic screening of the long-ranged Coulomb interaction in the
adiabatic limit.

The $\Delta_{1}/2$ anomaly involves momentum transfer between nodal
regions and could be important to understand the corresponding
self-energy corrections, in particular, the nodal kink observed in
ARPES experiments. Indeed, in a recent bulk-sensitive low photon energy
ARPES study \cite{Ref72} it has been demonstrated that electron-phonon
coupling is responsible for the nodal kink and the ''half-breathing''
mode $\Delta_{1}/2$ is the relevant mode.

Another general aspect becomes evident from our calculations namely
that the specific features of the solid-state chemistry of the cuprates
are essential. The distinctive ionic character of these compounds
underlines the relevance of the long-ranged polar Coulomb interaction
also in the doped metallic state. On the other hand, the Mott
insulating character of the undoped parent compounds being
antiferromagnetic insulators emphasizes the importance of the
short-ranged part of the interaction, notably by the repulsive on-site
interaction $U_{d}$. As discussed in this work the impact of both parts
of the interaction appears when studying the generic OBSM phonon
anomalies.

In the overwhelming number of attempts to describe the physics of the
cuprates the short-ranged part $U_{d}$ is considered exclusively and
simplified models with strongly reduced Hamiltonians are used. Such
models in the first place may be useful for the discussion of
theoretical ideas in general rather than for realistic calculations
where the material specifica must be included. Thus, at present it is
not at all clear if the two-dimensional one-band Hubbard model which is
most commonly applied for the cuprates contains sufficient information
to describe the real materials. In this context a recent combined DFT
and dynamical cluster Monte Carlo calculation for the three-band
Hubbard model is instructive and may serve as a guideline \cite{Ref73}
in the discussion of a minimal model for the cuprates. This study
demonstrates that the corresponding phase diagram of the model and in
particular a possible occurrence of superconductivity is quite
sensitive to the choice of hopping parameters and the down-folding
procedure used. In a recent study\cite{RefAimi} pairing correlations on
doped Hubbard models are reanalyzed using a sign-problem-free
Gaussian-Basis Monte Carlo method and it is concluded that the simple
Hubbard model does not account for high-temperature superconductivity.

The strong short-ranged repulsive $U_{d}$ by itself does not help with
the pair-binding, however,  favours $d$-wave symmetry and the usual
argument then is that as an implication of a large $U_{d}$ an
antiferromagnetic  exchange coupling $J$ arises which leads to an
attractive interaction between electrons of opposite spins on
neighbouring sites. $U_{d}$ is an unretarded particle-particle
interaction with no low-frequency dynamics and also the interaction
related to $J$ is unretarded.

On the other hand, several retarded attractive interactions mediated by
virtual bosonic excitations with low-frequency dynamics are discussed
in context with pairing in the cuprates. Most common are fluctuations
of spin and/or charge degrees of freedom of the electron liquid and
(adiabatic) phonons. From our findings in this paper low-frequency
coupled phonon-plasmon modes resulting from the strong polar
long-ranged interaction in the nonadiabtic region should be added to
the list. It still remains an open question which contribution and what
kind of cooperation between the different players is most important for
pairing and superconductivity in the cuprates. It seems that the high
as well as the low energy scale is essential for a synergetic interplay
of charge-spin- and lattice degrees of freedom which underlies the
physics in the cuprates in the normal as well as in the superconducting
state.

\section{SUMMARY AND CONCLUSIONS}
We have developed a realistic description of the electronic
bandstructure of LaCuO which is well suited to calculate the wavevector
and frequency dependent proper polarization part of the DRF for the
optimally to overdoped state. The latter is used to calculate the
adiabatic and nonadiabatic charge response and the coupled mode
dynamics in LaCuO.

The large anisotropy along the $c$-axis of the electronic structure of
the cuprates is considerably underestimated in DFT-LDA calculations. So
we have modified a LDA-based 31BM to account for the much weaker
interlayer coupling in the real material. We have optimized the
interlayer coupling by reducing some relevant hopping parameters
perpendicular to the CuO layer in such way that significant features of
the $\Lambda_{1}$ phonons polarized parallel to this axis are well
described. The $\Lambda_{1}$ modes have been chosen because they are
most sensitive in respect to the charge response orthogonal to the CuO
plane. The new sufficiently anisotropic model (M31BM) leads to a more
extended saddle point region, a reduction of the width of the
bandstructure, flatter bands in $c$-direction and a significant
narrowing and increase of the important van Hove peak. Simultaneously,
we obtain an associated amplification on the average of the proper
polarization part. These results demonstrate the importance of a
correct representation of the interlayer couplings being relevant for
both the electronic properties in the CuO layers and along the
$c$-axis.

Within the M31BM the measured Fermi surfaces for optimally and
overdoped LaCuO are well described. Clearly visible in the experiments
and the calculations is the change of the FS topology and of the
nesting structures upon doping. The latter are shown to be reflected in
the (noninteracting) susceptibility. Moreover, relevant Fermi surface
parameters for transport like the Drude plasma energy tensor and the
Fermi velocity tensor have been calculated and compared with standard
LAPW calculations. An enhancement of about a factor of 5 for the
anisotropy ratio is found for both types of parameters in the M31BM.

In our calculation of the phonon dispersion of the $c$-axis polarized
$\Lambda_{1}$ modes dramatic changes occur between the adiabatic limit
based on a static DRF and the nonadiabatic calculation founded on a
dynamic DRF. The former leads to static and the latter to dynamic
screening of the Coulomb interaction. On the other hand, virtually no
nonadiabatic effects have been detected for the phonons propagating in
the main symmetry direction $\Delta$ and  $\Sigma$ in the CuO plane.
This is due to the fast dynamics of the electrons in the CuO layer as
compared with the slow dynamics perpendicular to the plane. Only the
planar oxygen breathing mode $O^{X}_{B}$ experiences a minor
nonadiabatic correction worth mentioning which we attribute to
dynamical reduced nesting.

The adiabatic approximation as a basis paradigma to investigate lattice
vibrations fails severely in a small sector of $\vc{q}$-space around
the $\Lambda$ direction. The large nonadiabatic effect found for the
$\Lambda_{1}$ modes results from a low-lying $c$-axis plasmon predicted
within the M31BM. This collective mode couples via the strong
long-ranged polar Coulomb interaction to the optical phonons of allowed
symmetry. In particular we have shown that the large softening of the
O$_z^Z$ mode during the insulator-metal transition and its giant
linewidth found experimentally can be well understood in the
phonon-plasmon scenario.

At the $\Gamma$ point we identify ferroelectric-like $A_{2u}$ modes
with a large LO-TO splitting in form of plasmon-like and phonon-like
excitations. As seen in the experiments the ferroelectric-like $A_{2u}$
mode dominates the infrared response for polarization  along the
$c$-axis not only in the insulating state of LaCuO but also in the well
doped metallic state. Such an optical activity also found in our
nonadiabatic calculations for the metallic phase cannot be explained
using the adiabatic approximation with static screening for the
calculation of the phonon dispersion. This is what has been routinely
done in the literature in first principles calculations within static
DFT. So, the latter are not consistent with experimental evidence
concerning the $c$-axis response.

From our systematic investigation of the nonadiabatic charge response
and phonon dispersion we can identify a small region in $\vc{q}$-space
around the $\Lambda$ direction where the adiabatic approximation breaks
down. In this sector of $\vc{q}$-space the strong nonlocal,
nonadiabatic polar electron-phonon coupling leads to a mixed
phonon-plasmon dispersion. In order to experimentally resolve the
calculated dispersion a very high resolution perpendicular to the
$c$-axis would be needed. Presently, inelastic neutron scattering
experiments with such a high resolution are out of range and the actual
measurements only perform some average orthogonal to the $c$-axis where
the adiabatic charge response dominates by far because of the smallness
of the nonadiabatic zone.

We have calculated the induced charge response of the apex oxygen
stretching mode O$_z^Z$ in the adiabatic limit and the nonadiabatic
phonon-plasmon regime. Additionally we have computed the associated
orbital averaged changes of the potential felt by the electrons when
passing from the nonadiabatic phonon-plasmon region to the adiabatic
regime. In the adiabatic case we have an instantaneous interlayer
charge transfer while in the nonadiabatic situation the interlayer
charge transfer is dynamic in nature and phonon-like and plasmon-like
excitations are excited. Comparing with the adiabatic result we find
for the phonon-like O$_z^Z$ mode a strongly enhanced in phase charge
response which may be denoted as overscreening with respect to the
adiabatic case and accordingly a lower frequency appears while for the
plasmon-like mode with a higher frequency the charge response is
significantly weaker and runs out of phase, i.e. we observe an
antiscreening effect.

The related changes of the orbital averaged potential have been
calculated for the coupled phonon-plasmon modes. In particular for the
phonon-like mode we have found a very strong enhancment in the
nonadiabatic sector of phase space as compared with the adiabatic
result. The possible importance of the phonon-like O$_z^Z$ mode for the
electron self-energy has been pointed out. This demonstrates again the
importance of the long-ranged polar coupling of the electrons and the
$c$-axis phonons. As far as the low energy scale and the bosonic glue
to keep the electron pairs in the cuprates together is concerned our
calculations make a strong case that phonon-plasmon modes from the
nonadiabatic zone around the $c$-axis have to be added to the list of
possible players. Finally, it has been speculated within the
phonon-plasmon scenarium why SrRuO could be a low-temperature
superconductor in contrast to the high-temperature superconductor
LaCuO.

%%%%%%%%%%%%%%%%%%%%%%%%%%%%%%%%%%%%%%%%%
%% References
%%%%%%%%%%%%%%%%%%%%%%%%%%%%%%%%%%%%%%%%%

\end{document}